\documentclass[]{sn-jnl}

\usepackage[round]{natbib} 

\jyear{2023}%

\theoremstyle{thmstyleone}%
\theoremstyle{thmstyletwo}%

\theoremstyle{thmstylethree}%

\begin{document}

\title[Long-Term Evolution]{Long-Term Evolution of the Saturnian System}


\author[1]{\fnm{Matija} \sur{\' Cuk}}\email{mcuk@seti.org}

\author[2]{\fnm{Maryame} \sur{El Moutamid}}\email{maryame@astro.cornell.edu}

\author*[3]{\fnm{Giacomo} \sur{Lari}}\email{giacomo.lari@unipi.it}

\author[4,5]{\fnm{Marc} \sur{Neveu}}\email{marc.f.neveu@nasa.gov}

\author[6]{\fnm{Francis} \sur{Nimmo}}\email{fnimmo@ucsc.edu}

\author[7]{\fnm{Beno\^it} \sur{Noyelles}}\email{benoit.noyelles@univ-fcomte.fr}

\author[8]{\fnm{Alyssa} \sur{Rhoden}}\email{alyssa@boulder.swri.edu}

\author[9]{\fnm{Melaine} \sur{Saillenfest}}\email{melaine.saillenfest@obspm.fr}

\affil[1]{\orgdiv{Carl Sagan Center}, \orgname{SETI Institute}, \orgaddress{\street{339 N Bernardo Ave}, \city{Mountain View}, \postcode{94043}, \state{CA}, \country{USA}}}

\affil[2]{\orgdiv{Cornell Center of Astronomy and Planetary Sciences}, \orgname{Cornell University}, \orgaddress{\street{Space Sciences Building}, \city{Ithaca}, \postcode{14850}, \state{NY}, \country{USA}}}

\affil*[3]{\orgdiv{Dipartimento di Matematica}, \orgname{Università di Pisa}, \orgaddress{\street{Largo Bruno Pontecorvo 5}, \postcode{56127}, \city{Pisa}, \country{Italy}}}

\affil[4]{\orgdiv{Department of Astronomy}, \orgname{University of Maryland}, \orgaddress{\street{4296 Stadium Dr.}, \city{College Park}, \postcode{20742}, \state{MD}, \country{USA}}}

\affil[5]{\orgdiv{Planetary Environments Laboratory}, \orgname{NASA Goddard Space Flight Center}, \orgaddress{\street{8800 Greenbelt Rd.}, \city{Greenbelt}, \postcode{20771}, \state{MD}, \country{USA}}}

\affil[6]{\orgdiv{Department of Earth and Planetary Sciences}, \orgname{University of California Santa Cruz}, \orgaddress{\street{1156 High St}, \city{Santa Cruz}, \postcode{CA 95064},  \country{USA}}}

\affil[7]{\orgdiv{Institut UTINAM UMR 6213 / CNRS}, \orgname{Univ. of Franche-Comt\'e, OSU THETA}, \orgaddress{\street{BP 1615}, \postcode{25010}, \city{Besan\c{c}on Cedex}, \country{France}}}

\affil[8]{\orgdiv{Department of Space Studies}, \orgname{Southwest Research Institute - Boulder}, \orgaddress{\street{1050 Walnut St.}, \city{Boulder}, \postcode{80302}, \state{CO}, \country{USA}}}

\affil[9]{\orgdiv{IMCCE}, \orgname{Observatoire de Paris, PSL Research University, CNRS, Sorbonne Universit\'e, Universit\'e de Lille}, \orgaddress{\street{77 av. Denfert-Rochereau}, \postcode{75014}, \city{Paris}, \country{France}}}


\abstract{Here we present the current state of knowledge on the long-term evolution of Saturn's moon system due to tides within Saturn. First we provide some background on tidal evolution, orbital resonances and satellite tides. Then we address in detail some of the present and past orbital resonances between Saturn's moons (including the Enceladus-Dione and Titan-Hyperion resonances) and what they can tell us about the evolution of the system. We also present the current state of knowledge on the spin-axis dynamics of Saturn: we discuss arguments for a (past or current) secular resonance of Saturn's spin precession with planetary orbits, and explain the links of this resonance to the tidal evolution of Titan and a possible recent cataclysm in the Saturnian system. We also address how the moons' orbital evolution, including resonances, affects the evolution of their interiors. Finally, we summarize the state of knowledge about the Saturnian system's long-term evolution and discuss prospects for future progress.}




\maketitle


\section{Tidal Dissipation within the Planet and Orbital Evolution}\label{planet}


\par Satellites raise a tidal bulge on the planet they orbit. If there is any kind of lag in the planet's response, the tidal bulge will not point directly towards the satellite and as a result, the latter will experience a torque. Generally, planets spin faster than satellites orbit and, as a result, the torque pushes the satellites outwards. In the limit of low-eccentricity and inclination, a spin-synchronous satellite, and static solid tidal dissipation, for a tidal bulge averaged over an orbital cycle and apsidal precession, the semi-major axis increases as \citep{Murray-Dermott_1999}
\begin{equation}
    \dot{a}=3\frac{k_{2,p}}{Q_p}\sqrt{\frac{G}{M_p}}R_p^5m_s a^{-11/2}
    \label{eq:tidepla}
\end{equation}
where $G$ is the gravitational constant, $M_p$, $R_p$ and $k_{2,p}$ are the planet's mass, radius and tidal Love number, respectively, while $m_s$ and $a$ are the mass and semi-major axis of the satellite that raises the tides. Traditionally, the lag has been assumed to be proportional to $Q_p^{-1}$, where $Q_p$ is the so-called dissipation factor of the planet and is often taken to be constant. In this theory of ``equilibrium tides'', the semi-major axis evolves according to $t^{2/13}$, where $t$ is time \citep{Murray-Dermott_1999} and more distant satellites tend to evolve outwards more slowly. The result is converging orbits, which raises the possibility of capture into mean-motion resonance (see Section~\ref{mmr}). Tides on the planet also cause  the eccentricity of satellites to increase, even though this effect is generally smaller than the eccentricity damping due to the tidal dissipation within the moons (see e.g. \citealp{Goldreich-Gold_1963}).

More recently, a different mechanism, termed ``resonance locking'', has been proposed \citep{Fuller-etal_2016}. This mechanism is described in more detail in Chapter 2 of this volume. The key results, however, are as follows: 1) orbital migration can be significantly more rapid than previously thought; 2) outward evolution is exponential or power-law in time, rather than going as $t^{2/13}$, meaning that the rate of outwards motion accelerates; and 3) more distant satellites tend to evolve outwards faster. All three statements are true only if a satellite is in a resonance lock, which is not assured and depends on the internal structure of the planet.

Point (1) has received some observational confirmation from astrometry and radio science (\cite{Lainey-etal_2020} but cf. \cite{Jacobson_2022}). Point (2) means that an observed rapid present-day outward migration does {\em not} necessarily imply that the satellite formed recently. Point (3) means that the opportunities for capture into resonance are more limited than in the traditional picture, though the results are sensitive to very poorly known parameters.

Another important consequence of the resonance-locking hypothesis is that for satellites in a mean-motion resonance, the tidal heat production rate corresponding to constant (equilibrium) orbital eccentricities may be much higher than previously thought, because the effective $Q_p$ of the planet is lower \citep[e.g.][]{mey07}. This issue is discussed in more detail in Chapter~7 of this book\footnote{Chapter 7 has been published as \citet{nim23}}, and in Sections~\ref{moons} and \ref{enceladus} below.

\section{Mean-motion resonances }\label{mmr}

A mean-motion resonance (MMR) arises when two orbiting bodies (in this case satellites of a planet) have a ratio of orbital mean motions that is close to a simple fraction, i.e. 1/2, 2/3, 3/5 etc. MMRs are usually locations of strong gravitational interaction between the moons, and depending on many factors, this interaction could be temporary or the bodies may be permanently {\it captured into the resonance}, meaning that their period ratio stays fixed as their orbits evolve. 

More rigorously, two bodies are in resonance if a relevant angle called the {\it resonant argument} is librating rather than circulating. For a MMR, the general form of the resonant argument is $j_1 \lambda_1 + j_2 \lambda_2 + j_3 \varpi_1 + j_4 \varpi_2 + j_5 \Omega_1 + j_6 \Omega_2 $, where $\lambda$, $\varpi$ and $\Omega$ respectively designate the mean longitude, the longitude of pericenter and the longitude of the ascending node \citep{Murray-Dermott_1999}, and the subscripts 1 and 2 refer to the inner and the outer satellite. The coefficients $j_{1 - 6}$ are integers, with the condition that $j_1+j_2+j_3+j_4+j_5+j_6=0$ \citep[D'Alembert's rule, arising because the geometry of MMR does not depend on the origin of longitude;][]{Murray-Dermott_1999}. The order of the resonant argument is $\lvert j_3 \rvert+\lvert j_4 \rvert +\lvert j_5 \rvert +\lvert j_6 \rvert$, and the corresponding resonant term in the Hamiltonian (i.e. interaction potential) will be multiplied by $e_1^{\lvert j_3 \rvert} e_2^{\lvert j_4\rvert} s_1^{\lvert j_5 \rvert} s_2^{\lvert j_6\rvert}$, where $e$ and $s=\sin{(i/2)}$ refer to orbital eccentricity and inclination. As $e < 1$ and $s < 1$, and especially because the orbits of the major satellites of the giant planets tend to have small eccentricities and inclinations, lower-order resonances are typically stronger than higher-order ones. Note that each period commensurability ($j_2:j_1$) contains many formal resonances (often called ``sub-resonances") with different resonant arguments, which may affect either inclination or eccentricity. For example, within the 2:1 commensurability between Mimas (``M'') and Tethys (``$\Theta$''), we can have 1st order resonance (e.g. $2\lambda_{\Theta} - \lambda_M - \varpi_M$), 2nd order resonance ($4 \lambda_{\Theta} - 2\lambda_M - \Omega_M - \Omega_{\Theta}$, which is currently librating) as well as 3rd order resonance (e. g. $2\lambda_{\Theta} - \lambda_M + \varpi_M - \Omega_M - \Omega_{\Theta}$). In general, $\lvert j_1+j_2 \rvert$ is the lowest order that a resonance within the commensurability can have, but higher order sub-resonances are always also present. Note that $j_5+j_6$ has to be an even number \citep[due to inclinations being relative, rather than absolute, and the geometry of the MMR not depending on the choice of reference plane;][]{Murray-Dermott_1999}, with the implication that resonances affecting inclination have to be at least of second order, while eccentricity-affecting resonances can be of first order.     

Satellite systems of the giant planets introduce two elements in the resonant dynamics that are not as prominent for heliocentric orbits. Major satellites experience significant tidal torques from the primary, which makes moons migrate, potentially encountering resonances with each other (see Section~\ref{planet}). Additionally, the highly oblate shapes of the giant planets strongly perturb the orbits of the satellites, inducing fast orbital precession. Different sub-resonances of the same commensurability are separated in frequency space by secular (i.e. precessional) frequencies, so fast orbital precession leads to well-separated sub-resonances that do not overlap in frequency \cite[e.g.][]{MeyerWisdo:2008}. This non-overlap is important in order for a stable resonance capture to occur, as resonance overlap produces dynamical chaos \citep{Wisdom_1980}, which sometimes breaks resonances and sometimes leads to a much more complex evolution \citep[as seen in the Uranian system;][]{Dermo-etal:1988,Cuk2020}.

One fundamental rule of resonant encounters is that convergent orbital evolution between two bodies can lead to resonance capture, while divergent crossings of resonances always lead to ``jumps'', i.e. passages through resonance during which relevant orbital elements experience discontinuous ``kicks'' but no long-term resonant evolution \cite[e.g.][]{Dermo-etal:1988}. As the resonances within the same commensurability are separated by precession frequencies, it is always the case that inclination-type resonances (i.e. those with resonant arguments including longitudes of the node) are encountered first during convergent migration, as the precession of orbital nodes is generally retrograde (against the direction of orbital motion). Eccentricity-type resonances (which include the longitudes of pericenter in the resonant argument) tend to be crossed by converging satellites later as the pericenters precess in the direction of orbital motion. Whether convergent migration results in capture into a specific resonance depends on the parameters of the bodies and the resonance (lower-order resonances and higher masses make capture more likely), as well as migration rates (slower migration enhances capture probability) and pre-resonance orbital elements \citep[low eccentricity or inclination are required for capture;][]{Murray-Dermott_1999}. While dissipation within satellites does not directly affect the capture into resonances, it can indirectly enable capture by damping previously existing eccentricity (or inclination).



The above discussion assumes that a MMR is only affecting the two satellites whose orbits are in near-commensurability. Close proximity of several sizeable satellites in the Saturnian system means that three-body resonances are sometimes present. Most relevant three-body resonances are semi-secular ones involving Titan, in which the longitudes of the apse or node of Titan (the most massive moon and the most important perturber) appear in a resonant argument of a MMR between two moons other than Titan \citep{Cuk_ElMoutamid_2022}. Three-body resonances independent of two-body resonances are also possible, but tend to be relatively weak and any capture tends to be temporary \citep{Cuk_ElMoutamid_2022}. 


Once two moons are captured into a resonance, their semi-major axes evolve in unison in order to preserve the resonant relationship. In a stable MMR, eccentricity or inclination of at least one of the moons has to increase over time. This eccentricity or inclination is determined by the resonant argument, with the increasing orbital element being a conjugate of the secular angle in the resonant argument \citep{Murray-Dermott_1999}. For example, in the current Titan-Hyperion resonance with the resonant argument $4 \lambda_H - 3 \lambda_T - \varpi_H$ the secularly increasing element is $e_H$, conjugate to $\varpi_H$. Higher-order resonances can have orbital elements of both moons increase over time, as is the case with the inclinations of both Mimas and Tethys in the current Mimas-Tethys 4:2 resonance. The eccentricities and/or inclinations of bodies in MMR are expected to grow indefinitely, unless balanced by dissipation in the satellite or more usually until some other dynamical mechanism arrests further evolution or even breaks the resonance. Often this is due to the onset of resonance overlap. Even if closely-spaced sub-resonances are separate at the moment of resonant capture, their widths are proportional to eccentricities and inclinations and become wider over time. Eventually this can produce resonance overlap, resulting in dynamical chaos and likely breaking of the resonance relation.

One special case in which the orbital eccentricity can be constant indefinitely in the resonance is when the tidal dissipation within the satellite (described in the next subsection) acts to damp the eccentricity. Depending on the masses of the moons, their rates of tidal evolution and the strength of dissipation within the affected moon, the system can settle in an equilibrium in which the increase of eccentricity within the resonance is exactly counteracted by the damping by satellite tides \citep{mey07}. It is widely thought that Enceladus is currently in such equilibrium, in which eccentricity pumping by 2:1 MMR with Dione is counteracted by tidal dissipation within Enceladus, keeping eccentricity constant and providing energy for the geothermal activity on Enceladus \citep{lai12}. It is important to note that this situation can be stable only for some combinations of tidal dissipation parameters that damp the librations of the resonant argument, as in other cases the equilibrium may be temporary as librations grow and the moons evolve out of the resonance \citep{MeyerWisdo:2008}.

\section{Tidal dissipation and Heating within Satellites}\label{moons}


A synchronous satellite (i.e. one that has identical rotational and orbital periods) will experience a permanent tidal bulge as a consequence of its proximity to its parent planet. However, if the satellite's orbit is eccentric, the size and orientation of this tidal bulge will fluctuate; the latter is also true if the satellite has an axial tilt (obliquity). If there is any friction in the system, some of the mechanical energy deforming the satellite will be converted to thermal energy. This is the origin of tidal heating. 

Quantitatively, the heat production rate $\dot{E}$ is given by \cite[e.g.][]{Wisdo:2008}
\begin{equation}
    \dot{E}=\frac{3}{2} \frac{n^5 R^5}{G} \frac{k_2}{Q} \left(7e^2 + \sin^2\theta\right)
    \label{eq:Edot}
\end{equation}
Here $n$ is the satellite mean motion, $R$ is its radius, $e$ its eccentricity and $\theta$ its obliquity. This expression is correct in the limit of small $e$ and $\theta$. The quantity $k_2$ is the satellite's tidal Love number, which describes the response of its gravity field to the perturbing tidal potential; the maximum value (for a uniform, fluid body) is 1.5 and it is reduced if the body has appreciable rigidity or central concentration of mass. The quantity $Q$ describes the friction in the satellite; $Q^{-1}$ is proportional to the lag angle between the applied potential and the response, so that a low $Q$ implies a large lag angle and correspondingly higher dissipation.

Calculation of $k_2$ and (especially) $Q$ for satellites\footnote{Not to be confused with the analogous parameter $Q_p$ for the planet; see Section~\ref{planet}.} is not straightforward. In the simplest case, a solid viscoelastic (Maxwellian) body, the density, rigidity and viscosity at each point in the satellite need to be specified to derive $k_2/Q$ \citep[e.g.][]{Moore:2000}. Since viscosity is strongly temperature-dependent, and the heat production rate depends on $k_2/Q$, even this very simple case raises the possibility of feedbacks between heat production and internal structure \citep{OjakaSteve:1986,Hussm-etal:2004}. 

The Maxwell description of viscoelasticity, though often used, fails to adequately describe how real materials respond to forcing at different frequencies, so more complicated models (e.g. Andrade, Burgers) are needed \citep{renaud2018increased, renaud2021tidal}.
If oceans are present, other modes of tidal heating are possible, such as turbulent dissipation in the oceans themselves \citep{ChenNimmo2014}, or tidal flushing of water through a permeable crust \citep{Rovir-etal:2022}. However, the contribution of these mechanisms to the overall tidal heating budget appears in general to be small. 

Because tidal heating represents an energy loss, the semi-major axis of the satellite must contract. If it is not experiencing external torques, for angular momentum to be conserved the eccentricity must also decrease. Thus, for an isolated satellite any eccentricity will be damped relatively rapidly. The damping rate for small $e$ is given by \citep{Murray-Dermott_1999}
\begin{equation}
    \dot{e} = -\frac{a\dot{E}}{eGMm}
    \label{eq:edot}
\end{equation}
where $a$ is the semi-major axis, $m$ and $M$ the mass of the satellite and primary, respectively, and $\dot{E}$ is the tidal heating rate (equation~\ref{eq:Edot}).

If eccentricity tides dominate then Eq.~\eqref{eq:edot} can be used to derive the eccentricity damping timescale $\tau_e=e/\dot{e}$:
\begin{equation}
    \tau_e = \frac{GMme^2}{a\dot{E}}
    \label{eq:tauecc}
\end{equation}
Since $\dot{E}$ depends on $e^2$ (Eq.~\ref{eq:Edot}), the damping timescale is independent of eccentricity.

If dissipation is dominated by obliquity rather than eccentricity tides, the inclination ($i$) damping timescale $\tau_i$ for small inclination is given by \cite{chy89}
\begin{equation}
    \tau_i = \frac{GMm\sin^2{i}}{a \dot{E}}
\label{eq:tauinc}  
\end{equation}
Because the satellites are expected to be in a Cassini state\footnote{Cassini states are the end-points of energy damping within a rotating satellite, in which the obliquity of the satellite is determined solely by its shape and orbital inclination, and any initial free obliquity is damped \citep{Ward_1975}.}, the obliquity $\theta$ will scale with inclination and thus $\tau_i$ is independent of inclination. Other things being equal, inclination damping is slower than eccentricity damping because of the factor of $7 (\sin{i}/\sin{\theta})^2$ from Eqs.~\eqref{eq:Edot},\eqref{eq:tauecc} and \eqref{eq:tauinc}. 
Note that the factor $(\sin{i}/\sin{\theta})$ is typically $\gg 1$ for relatively close-in satellites of giant planets that are in the Cassini State 1 \citep{ChenNimmo2014}. However, for satellites with oceans, obliquity tidal heating can be much more effective than eccentricity tidal heating \citep{tyl08, ChenNimmo2014}, and so inclination damping may be more rapid than eccentricity damping.

Equation~\eqref{eq:edot} is the reason that orbital resonances (see Section~\ref{mmr}) are so important; although the eccentricity of an isolated satellite (like Triton) will swiftly damp to zero, if the eccentricity is being excited by an orbital resonance, then long-lasting tidal heating can arise. In this case the ultimate source of energy is the rotational kinetic energy of the planet. Because the damping rate depends on the tidal heating rate, and the tidal heating rate depends on the thermal structure, complicated and non-monotonic orbital and thermal histories can result \cite[e.g.][]{OjakaSteve:1986,Hussm-etal:2004}.

Because torques from the planet tend to increase the eccentricity and semi-major axis (Section~\ref{planet}), while dissipation in the satellite decreases both, for satellites in a MMR an equilibrium can result in which $e$ is constant (see Section~\ref{mmr}). \citet{mey07} show that the power of tidal heating of the inner moons of a resonant pair, assuming no tidal torque on the outer moon and all eccentricities being in equilibrium, is:
\begin{equation}
H = n_E T_E - {T_E \over L_E+L_D} \left( {G m_S m_E \over a_E} + {G m_S m_D \over a_D} \right) 
\label{meyer}
\end{equation}  
where $T_E$ is the tidal torque on Enceladus, while $L$ is angular momentum \textcolor{blue}{$(=ma^2n \sqrt{(1-e^2)})$}, $n$ is mean motion, $a$ is semi-major axis, $m$ is mass, and $G$ is gravitational constant (subscripts $E$, $D$ and $S$ refer to Enceladus, Dione and Saturn, respectively).  If we ignore terms of order $e^2$ \citep{mey07}:
\begin{equation}
H = n_E T_E  \left(1 - {1 + m_D a_E/(m_E a_D) \over 1+ (m_D/m_E) \sqrt{a_D / a_E}} \right) 
\label{meyer2}
\end{equation}
The torque on Enceladus is given by:
\begin{equation}
T_E={3 \over 2}{G m^2_E R^5_p k_{2p} \over a^6_E Q_p}
\label{torque}
\end{equation}
In this special case, the dissipation rate depends on the $Q_p$ of the planet but {\em not} $Q$ of the satellite, which affords a very significant simplification. Since the effective $Q_p$ of Saturn has been measured at various frequencies \citep{Lainey-etal_2020}, the equilibrium tidal heat production rate in satellites in MMRs (either current or ancient) can in principle be calculated \cite[e.g.][]{mey07}. Thus, for instance, equilibrium tidal heating rates of 11~GW and 4.8~GW are obtained for the 2:1 present-day Enceladus-Dione resonance and a putative earlier 3:2 Mimas-Enceladus resonance, assuming a Saturn $Q_p$ of 1,800 as suggested by astrometry \citep{Lainey-etal_2020}. A higher Saturn $Q_p$ would yield a correspondingly lower heat flow. Note that a tidal equilibrium for Enceladus requires it to be quite dissipative (with $k_{2E}/Q_E \simeq 0.01$ or even larger), likely due to the presence of an internal ocean, making solid-body dissipation estimates \citep[e.g.][]{Murray-Dermott_1999} invalid. Extensive further discussion of this issue is provided in Section \ref{enceladus} and Chapter~7.


\section{Important resonance passages in Saturn's moon system}\label{passages}

\subsection{Mimas-Tethys 4:2 MMR}\label{mimas}

\par Mimas and Tethys are locked in a 4:2 inclination-type MMR, associated with the resonant argument $2\lambda_M-4\lambda_\Theta+\Omega_M+\Omega_\Theta$. This argument slowly librates around $0^{\circ}$ with a large amplitude of $95^{\circ}$ and a period of 70 years. Such a resonance is expected to raise the inclinations of the two involved bodies over time. However, since Tethys is about 16 times more massive than Mimas, only the inclination of Mimas is significantly affected. \citet{Allan1969} derived the time evolution of these two inclinations since the trapping and has shown that, while the current inclination of Mimas ($1.6^{\circ}$) could be due to the excitation by this MMR, the one of Tethys ($1.1^{\circ}$) cannot, and was probably nearly as high as it is today when the capture happened. \citet{Cuk2016} propose an explanation for this inclination, involving a past 5:3 MMR between Dione and Rhea, followed immediately by a secular resonance between Tethys and Dione.

\par The trapping into this current resonance and its stability require that the orbits of Mimas and Tethys converge. Since the trapping, assuming no other perturbations, the libration amplitude has decreased from $180^{\circ}$ at the onset of resonance to the currently observed $95^{\circ}$, while the inclination of Mimas increased from the initial value to $1.6^{\circ}$. Using a linear model, \citet{Sinclair1972} found that when reaching the separatrix delimiting the resonance, the satellites had only a 4\% chance to be trapped. This low probability of the capture is due to the large current libration amplitude of the resonant argument, which implies an initial inclination of $0.4^{\circ}$ for Mimas. \citet{Champenois1999a,Champenois1999b} solved this paradox by showing that a non-linear model, which is more accurate, allows for secondary resonances to affect the libration amplitude in the main resonance. Allowing for much lower resonant libration amplitudes post-capture leads to much higher capture probabilities. These probabilities may almost reach 100\% for a null initial free inclination of Mimas, which means that when Mimas and Tethys encountered the resonance during their convergent migration, the trapping into the currently observed configuration was the only plausible outcome.

\par Before the current 4:2 inclination resonance was reached, Mimas and Tethys had to cross the sub-resonance with the argument $2\lambda_M-4\lambda_\Theta+2 \Omega_M$ (``$i_M^2$'' resonance). In the simplest case in which the pre-resonance inclination of Mimas was $0.4^{\circ}$ the probability of capture into the $i^2_M$ resonance was only 7\% \citep{luan14}, explaining why the capture did not happen. However, scenarios in which the initial inclination of Mimas was small and the libration amplitude of the current (``$i_Mi_{\Theta}$'') resonance was subsequently changed \citep[such as][]{Champenois1999b} would also make the capture into the $i_M^2$ resonance highly likely. Therefore any hypothesis on the origin of the current 4:2 $i_M i_{\Theta}$ Mimas-Tethys resonance would also need to account for the prior crossing of the $i_M^2$ resonance not resulting in capture.

\par None of these analytical and numerical studies actually constrain the dissipation rate inside Saturn, which is modelled as a scale temporal factor. The capture process and the evolution into resonance are assumed to be slow enough, i.e. adiabatic, to not be affected by the dissipation timescale.  In assuming a Love number $k_{2,p} = 0.341$ for Saturn and that the dissipation function $Q_S$ is constant over the whole orbital evolution, \citet{Champenois1999a} estimate that the capture happened some $17,000 ~Q_S$ years ago. This implies that the resonance was assembled only about 30~Myr ago using the $Q_S$ value from \citet{lai12}. Note that in equilibrium tidal theory (Eq. \ref{eq:tidepla}) orbital evolution timescales of Mimas and Tethys are relatively similar. {On the other hand}, if the tidal response of Saturn depends on the frequency \citep{Fuller-etal_2016}, the estimated age of the resonance could greatly differ from the value given above.  

\par Recent results, which imply non-equilibrium tides in Saturn, complicate the issue of the history and age of this resonance. Capture into resonance requires convergent evolution of Mimas and Tethys, which contradicts the end-member case in which all moon orbits are expanding with constant semi-major axis ratios \citep{Lainey-etal_2020}. While the directly measured tidal accelerations of Mimas and Tethys still have large uncertainties \citep{Lainey-etal_2020, Jacobson_2022}, the relatively fast observed evolution makes it unlikely that the age of the Mimas-Tethys resonance is greater than a few tens of Myr. Better determination of the moons' tidal accelerations is necessary to put firmer limits on the age of this resonance.

\subsection{Enceladus-Dione 2:1 MMR}\label{enceladus}

Enceladus and Dione are currently in a 2:1 MMR with an argument $2\lambda_D-\lambda_E-\varpi_E$ that acts to increase the eccentricity of Enceladus over time. The libration amplitude of this resonant argument is below one degree \citep{Murray-Dermott_1999}, implying long-term damping of the libration amplitude. Given that the eccentricity of Enceladus is modest ($e_E=0.0047$), it was always considered likely that the tides within Enceladus act to decrease this eccentricity (Section~\ref{moons}). This eccentricity damping releases heat within Enceladus, affecting its geophysics and geology. As the eccentricity of Enceladus is modified over time by satellite tides, it is not possible to use the current orbits of the moons to determine the age of this resonance.

After the {\it Voyager} mission, it was known that parts of Enceladus's surface were very young, but current tidal heating was considered insufficient for resurfacing due to low eccentricity and then-prevailing estimates of the rate of Enceladus's orbital evolution \citep{squ83}. As shown by \citet{mey07}, if we assume that the Dione-Enceladus resonance is in a long-term equilibrium, the resonance produces $(18,000/Q_S) \times 1.1$~GW of heat within Enceladus. $Q_S=18,000$ is the smallest value for which Mimas stays outside Saturn's Roche limit over the age of the Solar System (assuming frequency-independent equilibrium tides; Eq. \ref{eq:tidepla}). As it was widely accepted before the {\it Cassini} mission that Saturn's major moons were primordial, long-term average tidal heating of Enceladus was thought to be restricted to 1~GW or less.

{\it Cassini} observations have shown that the heat flux from the south pole of Enceladus is 10-15~GW \citep[][Chapter 6 of this book]{por06, how11}, an order of magnitude larger than predicted. Early solutions to this discrepancy suggested non-equilibrium scenarios \citep[e.g.][]{one10}, until \citet{lai12} found that fast tidal evolution of Saturn's moons suggested by astrometric observations can naturally explain the tidal heating in Enceladus being in equilibrium. Equilibrium heating of 15~GW implies $Q_E/k_{2E}=100$ for Enceladus (Section~\ref{moons}), giving us an eccentricity-damping timescale (in absence of excitation) of 0.5~Myr. Therefore Enceladus would need only a few Myr to settle into a dynamical steady state, and an assumption of equilibrium does not preclude geophysical evolution on longer timescales (Chapter 7 in this book). 

More recent findings of non-equilibrium, likely resonant, tides dominating the evolution of the Saturnian system \citep{Lainey-etal_2020} complicate this picture. Just like with the Mimas-Tethys resonance (previous subsection), the existence of the Enceladus-Dione resonance requires that the orbits of those two moons converge. Furthermore, unlike in the case of Mimas-Tethys resonance which only requires past net convergent evolution, tidal heating within the Enceladus-Dione resonance requires ongoing orbital convergence. This is consistent with the small libration amplitude which would be growing if the two moons' orbits were diverging, making the resonance evolve toward dissolution.

Convergent evolution is not compatible with the end-member orbital evolution model in which all moons are locked to resonant modes that all evolve on the same timescale \citep{Lainey-etal_2020}, or with a model in which the modes are divergent \citep{Fuller-etal_2016}. However, a situation in which Enceladus is in a resonance lock, but Dione is not, can result in a MMR. Using the approach of \citet{mey07} and adding the assumption about Enceladus being much less massive than Dione, we can roughly estimate the (equilibrium) tidal heating of Enceladus in the resonance as:
\begin{equation}
H \approx {\frac {G m_S m_D}{2 t_a a_D}} = {\frac {5 \times 10^{28} \ {\rm J}} {t_a}} 
\label{heat}
\end{equation}
where $t_a=(a_D/\dot{a}_D)_{MMR}$ is the timescale for the evolution of Dione's orbit forced by resonance with Enceladus. In the case of equilibrium tides, $t_a$ would be the timescale of the two moons' orbital convergence multiplied by the moons' mass ratio $m_D/m_E$, so $t_a \approx 100$~Gyr if $Q_S=1700$, yielding $H \approx 16$ GW. If Enceladus is evolving through a resonant lock, then $t_a$ is simply the timescale of the resonant mode's evolution, as Dione would be forced to evolve at the same rate.  If we assume $t_a=9$~Gyr, as proposed by \citet{Lainey-etal_2020} as being a typical value for orbital evolution due to resonance locking in the Saturnian system, $H \approx 180$~GW, more than an order of magnitude in excess of the observed value \citep{how11}. 


On the other hand, if the resonance lock evolves uniformly in a reference frame rotating with Saturn and the satellites' orbits evolve divergently \citep{Fuller-etal_2016}, steady-state heating of Enceladus will be much lower. The Saturn-frame resonance lock would produce $a/\dot{a} \approx 200$~Gyr for Enceladus, based on Titan's evolution timescale of 11~Gyr \citep{Lainey-etal_2020} and $a/\dot{a} \propto {n}/(\Omega_p-n)$, where $\Omega_p$ is the planet's rotation rate \citep{Fuller-etal_2016}. This rate of evolution gives us a steady-state heat flow of about 8~GW for Enceladus, relatively close to the measured value, but almost certainly less than the true heat production rate. However, if we use Rhea's observed evolution \citep[$a/\dot{a}=6$~Gyr;][]{Lainey-etal_2020} to estimate Enceladus's tidal heating in the same model, we get $t_a \simeq 27$~Gyr and $H \approx 60$~GW. This last estimate is motivated by the main result of \citet{Jacobson_2022}, who finds that Titan does not exhibit non-equilibrium tidal evolution but Rhea does. If we use the observational results for the secular acceleration of Rhea from \citet{Jacobson_2022}, the timescale for Enceladus's evolution through resonant lock (assuming divergent modes) could be in the 30-40~Gyr range and the associated tidal heating in the 40-55~GW range. Note that the total tidal heating rate of Enceladus must exceed the measured value, as the thermal observations are not sensitive to distributed tidal heating outside the South Polar Terrain. Models of Enceladus's ice shell suggest a heat loss rate in the range 25-40~GW  \citep{hemingway2019enceladus}, so there may not be a discrepancy between the heating rates predicted here and the actual heat production rate (see Chapter~7).

A separate constraint on the possible resonant-lock driven heating of Enceladus comes from the observational limits on the orbital deceleration of Dione. The minimum possible effective $k_{2,p}/Q_p$ for Dione found by \citet{Lainey-etal_2020} is $0.6 \times 10^{-4}$, which implies a slowest evolution of about 30~Gyr. This means that 55~GW is the minimum tidal heating \textcolor{blue}{in the Enceladus-Dione system} allowed by the astrometric observations, if all of Dione's orbital evolution is due to Enceladus. \citet{Jacobson_2022} has larger error bars on the observed evolution of Dione, and allows for minimum heating rates that are about twice smaller. Note that these ``lower limits'' apply only to pure evolution of Enceladus by resonance lock that includes zero tidal torque on Dione. Models in which equilibrium tides are significant are not constrained by these limits.

We conclude that Enceladus's observed heat flux (which is a lower bound on the actual heat production rate) may be consistent with the resonance lock model in which the resonant modes are divergent, but is well short of that expected from the simple picture in which all resonant modes evolve on the same timescale. A picture in which modes affecting the inner satellites evolve more slowly \citep{Fuller-etal_2016} can explain the current Enceladus-Dione resonance, but requires Dione to have avoided direct resonance locking to an internal mode within Saturn. Further refinement of our understanding of Enceladus's tidal heating will require more precise measurement of the current orbital evolution of the Saturnian moons, as well as more complete constraints on the thermal flux from Enceladus (Chapter 6). Further discussion of Enceladus's tidal heat budget and thermal evolution may be found in Chapter~7.

Unlike in the case of the Mimas-Tethys resonance, less attention has been given to the formation of the Dione-Enceladus resonance than to its continuing maintenance. In order for Enceladus and Dione to reach the current sub-resonance of their 2:1 MMR through convergent orbital evolution, numerous other sub-resonances must be crossed first. Past work \citep{MeyerWisdo:2008, zha09} has concentrated on the sub-resonances that can be modelled assuming two bodies on planar orbits, including the second-order 4:2 $e_E e_D$ mixed-eccentricity resonance. However, inclination-type resonances (equivalent to the one Mimas and Tethys are currently in) would be encountered first, and their effect on the (currently very low) inclination of Enceladus has yet to be fully explored. Additionally, using numerical integrations featuring all major moons, \citet{Cuk_ElMoutamid_2022} have recently shown that two-body resonances also contain additional three-body sub-resonances that can result in capture. Clearly more work is needed to understand the history of the Enceladus-Dione 2:1 resonance.

\subsection{Past Resonances Between Mid-Sized Moons?}\label{dione}

\begin{figure}[t]
\centering
\includegraphics[scale=.4]{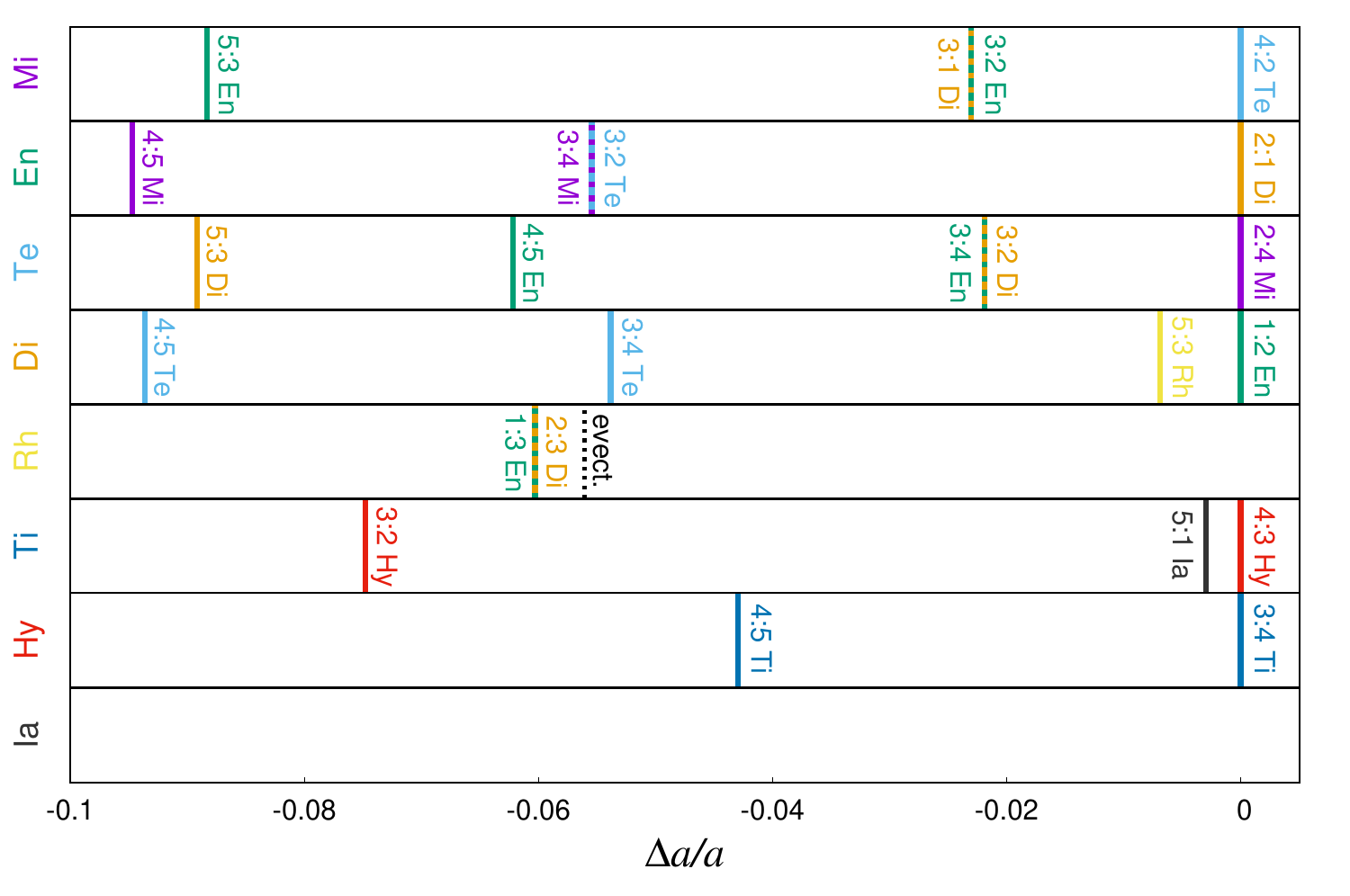}
\centering
\caption{Current location of nominal MMR for each satellite in terms of variation of its semi-major axis, assuming that the other moons do not move. Within a $10\%$ variation in $a$, we reported all first-order resonances up to 5:4, all second-order resonances up to 5:3, the evection resonance for Rhea and the 5:1 resonance between Titan and Iapetus. Overlapping between resonances is due to currently active MMR.}
\label{reslocations}
\end{figure}


In addition to currently active resonances, researchers have long suspected that Saturn's moons may have experienced additional resonances in the past which have since been broken (Figure \ref{reslocations}). Here we will address several proposed past resonances among the inner moons (out to Rhea), while those involving Titan and the Sun (i.e. the evection resonance) are discussed in separate subsections. For an introduction to the topics of orbital migration, resonance capture and evolution within the resonance, we refer the reader to Section \ref{mmr}.

Among the inner moons, past resonances are often considered responsible for excited eccentricities and inclinations that cannot be accounted for with observed interactions. For example, while the inclination of Mimas is likely a product of its current 4:2 MMR with Tethys, the eccentricity of Mimas is unaffected by this resonance and must predate it. Similarly, the large inclination of Tethys mostly precedes the resonance with Mimas, requiring an origin mechanism (Section~\ref{mimas}). Here it is assumed that the moons formed on coplanar, circular orbits, as both eccentricity and inclination are damped by expected collisions with satellitesimals \citep{pea99}. Apart from current orbital elements, the record of tectonic activity, such as Ithaca Chasma on Tethys, could indicate past tidal heating driven by high eccentricity \citep{ChenNimmo2008}. Tethys's orbital eccentricity is presently very low, but past high eccentricity implies that Tethys experienced one or more orbital resonances during its history. 

Reconstruction of past resonance crossings among the inner moons is highly dependent on the assumed strength and nature of tidal dissipation within Saturn. The potential past resonances that were studied in past work and are therefore discussed here are expected to have happened in the context of equilibrium tides. Equilibrium tidal theory predicts that all adjacent inner moon pairs are on converging paths, with the sole exception of Enceladus and Tethys, which were thought to be on divergent orbits \citep{Murray-Dermott_1999}. Frequency-dependent tidal dissipation within Saturn certainly changes this picture and may rule out some past resonances while introducing new ones. Here we will review  past work on these resonances and follow with a brief discussion on how the latest results on Saturn's tidal response \citep{Lainey-etal_2020, Jacobson_2022} change our assessment of these proposed resonant encounters.

\

{\bf Mimas-Enceladus 3:2 Resonance:} In the classical picture of equilibrium tides, one of the most recent resonance passages was that of the Mimas-Enceladus 3:2 MMR. This is expected to have been a convergent encounter, and has been proposed by \citet{MeyerWisdo:2008} as the source of Mimas's eccentricity. In particular, \citet{MeyerWisdo:2008} have suggested capture into $e_M$ and $e_M e_E$ (i.e. the 6:4 Mimas-Enceladus MMR) as possible sources of Mimas's eccentricity, while ruling out past capture into the $e_E$ sub-resonance as it would have been impossible to break. \citet{TianNimmo2020} later revisited this work using a wider range of tidal parameters, and found regions of tidal parameter space in which the hypothesis of \citet{MeyerWisdo:2008} is likely. Both of these studies relied on semi-analytical treatment of the resonant Hamiltonian, and assumed a planar system. The last issue is crucial, as the inclination of Enceladus is presently very low ($i_E \approx 0.01^{\circ}$), putting strong constraints on past resonance crossings involving Enceladus. \cite{ElMoutamid2019} 
proposed that a long-lived Mimas-Enceladus 3:2 $e_E$ resonance was broken when the pair encountered respective 3:1 and 2:1 resonances with Dione. However, ongoing work by some of this chapter's authors (El Moutamid and \'Cuk) using direct numerical integrations suggests that this triple resonance usually produces dynamical chaos, exciting the inclination of Enceladus beyond the observed value. In principle, a high inclination could have been damped by fluid tides in Enceladus's ocean (Section~\ref{moons}) but no quantitative study has been attempted. Clearly more work is needed to explore the dynamical viability of a past Mimas-Enceladus 3:2 MMR crossing, especially as the tidal evolution rates of these two moons (which may also not be constant over time) are hard to constrain by other means.

\
{\bf Mimas-Dione 3:1 Resonance:} The fact that Enceladus and Dione are currently in a 2:1 resonance implies that their orbital evolution (at least recently) has been convergent. Therefore, before this resonance was established Enceladus must have been interior to the resonant location. A clear implication is that if Mimas ever crossed the 3:2 MMR with Enceladus, it must have also crossed the 3:1 resonance with Dione (either at the same time or later). The reverse is not true: depending on the relative tidal evolution rates and the age of the system, Mimas could have in principle crossed the 3:1 resonance with Dione and not the 3:2 MMR with Enceladus. Given the large mass difference between Dione and Mimas, Mimas is likely to be affected by their 3:1 MMR much more than Dione. \citet{MeyerWisdo:2008} suggested the Mimas-Dione 3:1 MMR as another possible source of the eccentricity of Mimas. They prefer capture into the $e_M e_D$ sub-resonance as it would also excite the eccentricity of Dione in line with constraints from Dione's later resonance passage with Enceladus. More recently \citet{Cuk_ElMoutamid_2022} simulated the Mimas-Dione 3:1 resonance and found that three-body resonances involving the eccentricity of Titan can be important. This three-body sub-resonance, not included in the semi-analytical model of \citet{MeyerWisdo:2008}, usually leads to a short-lived capture that would excite the eccentricity of Mimas, consistent with the present state. Since Mimas's eccentricity is not being excited at present, survival of its eccentricity from an earlier time places limits on how dissipative it could be. At present, the 3:1 Mimas-Dione resonance appears to be a viable candidate for the source of Mimas's eccentricity, but more work is needed to confirm this. 

\

{\bf Tethys-Dione 3:2 Resonance:} This resonance should have taken place relatively recently if Tethys and Dione were to follow equilibrium tidal theory in their orbital evolution. In the classical picture in which Saturn's tidal $Q_S=18,000$ \citep{Murray-Dermott_1999} and is constant, this resonance would have happened about 1~Gyr ago, while $Q_S \approx 1,800$ proposed by \citet{lai12} would put this resonance at 100~Myr ago. This resonance was proposed by \citet{ChenNimmo2008} as the source of tidal heating of Tethys, and this hypothesis was explored in more detail by \citet{ZhangNimmo2012} using numerical integrations. \citet{ZhangNimmo2012} found that the Tethys-Dione 3:2 resonance was impossible to break once established, and proposed a large impact on Tethys (that formed the Odysseus basin) as a mechanism of breaking this resonance. However, \citet{ZhangNimmo2012} assumed  a planar system, therefore ignoring all constraints from the satellites' inclinations, and also underestimating the amount of chaos in the system (as all inclination-type resonances were absent in their simulations). \citet{Cuk2016} studied this resonance using direct integrations with non-zero inclinations, and found that the inclination of Tethys is excited up to several degrees, after which a chaotic phase excites both the eccentricities and inclinations of both moons. The chaotic interactions ultimately break the resonance, but leave Tethys and Dione with inclinations well in excess of the observed ones. Unless both bodies experienced efficient inclination damping, these results imply that the Tethys-Dione 3:2 MMR never happened and therefore that the present inner satellite system of Saturn must be younger than 100~Myr \citep[][assuming $Q_p$=1,800]{Cuk2016}. However, the recent astrometric results implying that the tidal response of Saturn is frequency-dependent suggest that Saturn's response to Tethys is significantly {\it weaker} than to other moons \citep{Lainey-etal_2020, Jacobson_2022}. A slowly-migrating Tethys would have either converged with Dione much more slowly or not at all (observations are still too uncertain to distinguish between these possibilities), so this resonance clearly cannot be used to put an upper limit on the age of the system. In summary, a passage through the Tethys-Dione 3:2 resonance would have produced inclinations inconsistent with the present values \citep[][unless the inclinations of both Tethys and Dione were greatly damped by some process]{Cuk2016}, and its past occurrence is also disfavored by recent observational results on the moon's orbital evolution \citep{Lainey-etal_2020, Jacobson_2022}, so this resonance is unlikely to have much bearing on the history of the system.

\

{\bf Dione-Rhea 5:3 Resonance:} In the equilibrium tides-based relative chronology, the Dione-Rhea 5:3 resonance happens soon after the Tethys-Dione 3:2 resonance. \citet{Cuk2016} studied the former using numerical simulations and unexpectedly found that this resonance between Dione and Rhea is a good candidate for the origin of Tethys's inclination. They find that following a largely chaotic Dione-Rhea 5:3 MMR passage, Dione and Tethys encounter a secular resonance with the argument $(\varpi_{\Theta} + \Omega_{\Theta}) - (\varpi_D - \Omega_D)$ (see Section \ref{mmr} for definitions). The two expressions within parentheses designate very slow-moving angles, as the apsidal and nodal precession rates for each moon are approximately equal and in opposite directions. However, precession rates are perturbed in proximity of MMRs, and in this case Rhea's perturbations just outside the Dione-Rhea 5:3 MMR modify the sum of Dione's apsidal and nodal precession rates just enough to produce the secular resonance. In the secular resonance, the eccentricity and inclination of Dione are transferred to Tethys in equal amounts (i.e. $\sin(i_{\Theta}) \approx e_{\Theta}$). This sequence of resonances (Dione-Rhea 5:3 MMR followed by Tethys-Dione secular resonance) can explain the high inclinations of Tethys ($i_{\Theta} = 1^{\circ}$ before resonance with Mimas) and Rhea ($i_R=0.33^{\circ}$) despite the low inclination of Dione ($i_D=0.02^{\circ}$, decreased by the secular resonance). These resonances also produce past large eccentricity of Tethys ($e_{\Theta} \simeq 0.01$) and the current eccentricity of Dione, assuming reasonable tidal dissipation. A past excited eccentricity of Tethys could in principle explain a past heating episode on Tethys inferred from Ithaca Chasma \citep{ChenNimmo2008}. The free eccentricity of Rhea, however, is currently just $e_R \approx 2 \times 10^{-4}$ requiring either strong dissipation or some other mechanism for damping. 


\par The biggest challenge to the past passage through the Dione-Rhea 5:3 MMR is the currently observed tidal evolution of Rhea \citep{Lainey-etal_2020}, which is five times faster than equilibrium tides would predict and would make the orbits of Dione and Rhea divergent. Two obvious solutions are that either the inclinations of Tethys and Rhea require different explanations, or Rhea's fast evolution is a phenomenon that started relatively recently. More work is urgently needed, both on exploring alternative possible sources of Tethys's inclination, and on the nature and operation of tidal dissipation within Saturn. 

\subsection{Rhea and the Evection Resonance}\label{rhea}


\par The evection resonance is a semi-secular resonance in which a moon's precession period is equal to the parent planet's orbital period \citep{tou98}. While no moons in the solar system are currently in the evection resonance, studies of the early orbital history of Earth's Moon have identified the evection resonance as an important dynamical mechanism \citep{tou98,cuk12, tia17, ruf20}. Evection is a ``semi-secular'' resonance in the sense that the (apparent) mean motion of the perturber (the Sun) is in resonance with the apsidal precession of the satellite's orbit. As the moons' orbital precession are overwhelmingly driven by Saturn's oblateness, the dynamics of the evection resonance does not depend on the relative arrangement of moons, as MMR do. The evection resonance affects any moon at a certain distance from Saturn, which is given by:
\begin{equation}
a_{eve} = R_S \left( {\frac 3 2} J_2 {\frac {\Omega_0} {n_s}} \right) ^{2/7} 
\label{eve}
\end{equation}
where $J_2$ is Saturn's oblateness moment, $\Omega_0 =\sqrt{G M / R_S^3}$ is the orbital frequency at $a=R_S$ (Saturn's radius), and $n_S$ is Saturn's heliocentric orbital mean motion. Equation~\eqref{eve} gives $a_{eve}=8.1 R_S$, but Titan's perturbations shift the location of the evection resonance to $a_{eve}=8.2 R_S$. This distance is somewhat smaller than the current orbital distance of Rhea (which has $a=8.7 R_S$). 

\par Estimates of the past orbital evolution of Rhea based on extrapolation from observations \citep{Lainey-etal_2020, Jacobson_2022} suggest that Rhea should have crossed the evection resonance about 300-400 Myr ago. The evection resonance, with the resonant argument $2 \lambda_S - 2\varpi$ ($\lambda_S$ is Saturn's mean longitude, $\varpi$ Rhea's longitude of pericenter), is expected to significantly excite Rhea's eccentricity. One possible outcome is that Rhea's satellite tides become strong enough to push the satellite inwards through the resonant mode, leaving it stranded interior to the mode. Another possibility would be Rhea evolving with the mode, while captured in the evection resonance, producing a continuously increasing eccentricity. Yet another outcome would be a Rhea that continues evolving through the mode, but moves beyond the resonance, with Rhea's excited  eccentricity slowly damping. Only the last of these possibilities would be consistent with the present orbital distance, small eccentricity ($\approx 0.001$) and observed evolution rate of Rhea.

\par
\begin{figure}[ht]

\centering
\includegraphics[scale=.7]{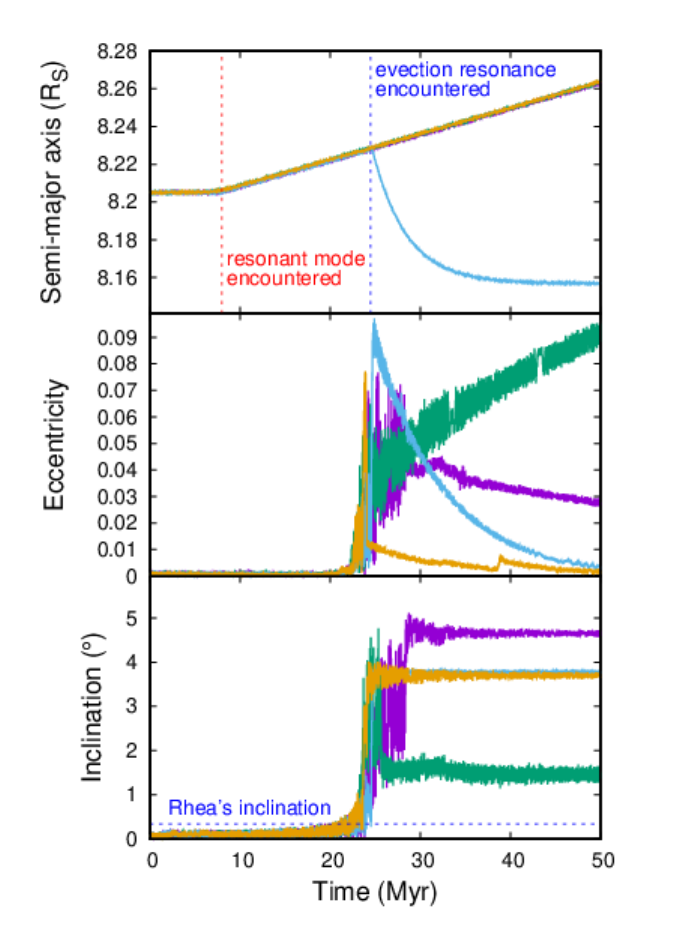}
\centering
\caption{Simulations of Rhea's encounter with the solar evection resonance using different parameters for the width of Saturn's resonant mode and Rhea's tidal dissipation. From top to bottom, the panels plot Rhea's semi-major axis, eccentricity and inclination. The purple and the green lines had $Q/k_2=1000$ for Rhea, and the light blue and the orange had $Q/k_2=100$ (i.e. likely value for Enceladus). The purple and the light blue lines had effective tidal $Q = Q_0 + \left({\frac {n - \nu}{\nu \sigma}} \right)^2$ (where $\nu$ is the mode's frequency and $\sigma=10^{-6}$) due to the resonant mode, while the green and the orange lines used the same $Q$ profile with $\sigma=10^{-5}$. The evolution timescale is $a/\dot{a}=6$~Gyr, as found for Rhea by \citet{Lainey-etal_2020}.The dotted line on the bottom panel shows the present-day inclination of Rhea. This simulation included the Sun, Jupiter, Saturn, Titan, and Rhea. From work in preparation by M\'C. \label{evection}} 
\end{figure}

 Fig. \ref{evection} shows four preliminary simulations done by author M.\'C. using a modified version of the numerical integrator {\sc simpl} \citep{Cuk2016}. One of the simulations shows a long-term resonance capture (green line), one ``drops out'' from the resonant mode (light blue line), while the other two simulations have Rhea staying with the mode but moving beyond the evection resonance. However, the bottom panel that plots inclination, indicates that the evection and associated resonances also significantly excite the {\it inclination} of Rhea, well in excess of the observed value ($i_R=0.33^{\circ}$).

\par The excitation of inclination appears to be chaotic and occurs concurrently with the evection resonance. This is consistent with results obtained by \citet{Cuk2016} assuming that Rhea migrated by equilibrium tides. The best candidate for the excitement of the inclination is the resonance as $\lambda_S-\varpi_S+\Omega-\Omega_S$ where $\Omega$ refers to the longitude of the ascending node and the subscript $S$ refers to Saturn's heliocentric orbit. As the reference plane is the Laplace plane\footnote{See Section~\ref{ssec:ob-basics} for a definition.} of Rhea's orbit, which is close to Saturn's equatorial plane, $\Omega_S$ is identical to the longitude of Saturn's equinox. Both $\Omega_S$ and $\varpi_S$ are relatively slow-precessing (with periods of about 1.8~Myr and 50 kyr), so the resonance happens because $\dot{\lambda}_S = n_S \approx -\dot{\Omega}$. One can think of this resonant term as a combination of the Sun's ``annual equation'' \citep[a term in classical lunar theory;][]{bro61} and the Sun's main secular perturbation on the inclination of Rhea. In other words, the Sun induces perturbations in Rhea's inclination once per precession period, and solar perturbations vary in strength during one orbit of Saturn because of the planet's orbital eccentricity. When these two cycles are commensurable, this strong resonant perturbation occurs. This resonance is chaotic, as its strength depends on the eccentricity of Saturn's orbit, which varies in the $0.01 < e_S < 0.09$ range with a period of $5 \times 10^{4}$~yr due to Jupiter's perturbations. Jumps in inclination also affect the apsidal precession, making the main evection resonance also chaotic. Direct numerical integrations are clearly the only practical and accurate way of studying this problem.

\par In some cases, the orbital inclination of a moon can be damped by obliquity tides \citep{chy89}, but this is unlikely to be applicable to Rhea. Large-scale inclination damping requires large forced obliquities, either through a Cassini State transition like the one experienced by the Moon \citep{ChenNimmo2016}, or spin-orbit resonance, as proposed for Uranian moons \citep{Cuk2020}. The forced obliquity of Rhea is currently $\theta=0.03^{\circ}$ \citep{ChenNimmo2014}, and the timescale for inclination damping is longer than that for eccentricity damping by a factor of $7 (\sin{\theta}/\sin{i})^2$ \citep{chy89}, which in the case of Rhea is on the order of $10^3$. If Rhea is as dissipative as ``Enceladus in equilibrium'' with $Q/k_2=100$, the eccentricity damping timescale is on the order of 10~Myr (Fig. \ref{evection}), which puts the timescale for inclination damping to many Gyr. Resonant tides excited in satellite oceans offer a different mechanism of inclination damping \citep{tyl08}, but \citet{ChenNimmo2014} find that ocean obliquity tides would produce less dissipation in Rhea than classic obliquity tides that assume $Q/k_2 \simeq 100$. The above upper limits on dissipation may be overestimates, as Rhea currently does not appear to be in hydrostatic equilibrium \citep{tortora2016rhea}, and a solid tidal Love number $k_2=0.01$ \citep{ChenNimmo2014} may be more applicable. Therefore, it appears that the damping of multi-degree inclination of Rhea within the last 0.5 Gyr is implausible.

\par It appears that Rhea should have crossed the evection resonance about 300-400 Myr ago and acquired a large inclination that could not have been subsequently damped. This dynamical history is clearly in conflict with Rhea's modest inclination ($i_R=0.33^{\circ}$), and this disagreement suggests that one or more assumptions that produced this result are incorrect. While it cannot be completely ruled out that Rhea somehow avoided acquiring high inclination during resonance crossing, numerical results available so far suggest that it is highly unlikely that Rhea crossed the evection resonance, at least in the present dynamical environment. If the obliquity of Saturn is only 1-2~Gyr old \citep{Saillenfest-etal_2021a}, it is possible that Rhea did cross the evection resonance in the distant past when the out-of-plane perturbations by the Sun were much weaker. This would still require a separate solution for the timescale problem. One possible explanation is that Rhea became locked to a normal mode \citep{Fuller-etal_2016} more recently than 400~Myr ago, with its orbital evolution before that being much slower. Another explanation would be that Rhea re-accreted outside the evection resonance distance, as a part of a dynamical cataclysm \citep[e.g. ][]{Cuk2016}. In order to distinguish between these possibilities a better understanding of tidal dissipation within Saturn, both past and present, is likely needed.

\par The only firm conclusions we can draw from the proximity of Rhea to the evection resonance is that we can exclude the global picture proposed by \citet{Lainey-etal_2020} in which the moons evolve by resonance lock practically since the system's formation.  Unlike the tidal evolution of Titan where the results of \citet{Lainey-etal_2020} and \citet{Jacobson_2022} are in disagreement, both studies agree on the fast orbital evolution of Rhea, corresponding to a timescale of $(a/\dot{a}) \approx 6$~Gyr. Large-scale migration of Rhea, if it is a long-term steady state, assures relatively recent (300-400~Myr) crossing of the evection resonance which would overexcite Rhea's inclination. Therefore, within the last few hundred Myr, Rhea must have experienced either a dramatic speeding up of its tidal evolution, or possibly a disruption and re-accretion.


\subsection{Titan-Hyperion 4:3 MMR}\label{hyperion}


\par Titan and Hyperion are currently in the 4:3 MMR, which keeps their orbital periods fixed in a 3:4 ratio. As Hyperion's orbit is relatively eccentric ($e_H=0.104$), the resonance is important for protecting Hyperion from having close encounters and colliding with Titan. It is established that the eccentricity of Hyperion in the resonance grows as Titan migrates outward \citep{Murray-Dermott_1999}; in the context of equilibrium tides, the orbital evolution of Hyperion on its own has a negligible effect on the resonance. Therefore, the eccentricity of Hyperion could be used to constrain the past orbital evolution of Titan. 

\par According to Eq. 8.242 in \citet{Murray-Dermott_1999}, the increase in the eccentricity of Hyperion is given by:
\begin{equation} 
{\frac {\dot{e}_H}{e_H}} = {\frac 1 {e_H^2}} {\frac {m_T}{m_S}} n_H a_H {\frac F {3 g}}
\label{md1}
\end{equation}
where $m$, $a$, $e$ and $n$ are masses, semi-major axes, eccentricities and mean motions as defined before, and subscripts $S$, $T$, and $H$ refer to Saturn, Titan and Hyperion. The other variables are defined as $F = 4 \dot{n}_H - 3 \dot{n}_T \approx -3 \dot{n}_T$, and $g=16 G m_T/a_H^2+9 G m_H/a_T^2 \approx 16 G m_T/a_H^2$. Expanding $F$ and $g$ into Eq.~\eqref{md1}, and using $G m_S/a_H^3 = n_H^2$, $n_H=(3/4)n_T$ and $\dot{n}/n=-(2/3)\dot{a}/a$, we get:
\begin{equation}
2 e_H \dot{e}_H = {\frac {{\rm d} (e^2)} {{\rm d}t}} = {\frac 1 4} {\frac {\dot{a}_T}{a_T}}
\label{md2}
\end{equation}
Assuming that Hyperion started with a low eccentricity, and evolved to $e_H=0.1$, Titan's semi-major axis must have evolved by about 4\%, assuming a constant rate of migration. This is a very rough estimate, but it gives us the correct order of magnitude of how much Titan could have migrated since Hyperion was captured into the resonance. 

\par Before 2012, when Saturn was estimated to have $Q_p>18,000$ \citep{Murray-Dermott_1999}, Titan was thought to have migrated less than 1\% of its orbital distance, and therefore the resonance was thought to have been established and largely evolved by non-tidal means, such as gas drag in protosatellite nebula \citep{pea99}. \citet{gre73} proposed an alternative view that the Titan-Hyperion resonance was evolved by tides alone, and implied a much faster global evolution rate of Saturn's moons.

\par When \citet{lai12} presented evidence for the fast tidal evolution of Saturn's moons, \citet{Cuk-etal_2013} noted that the new value of Saturn's tidal quality factor $Q_p \approx 1,700$ would make the Titan-Hyperion resonance about as old as the Solar System. The more recent direct measurement of Titan's orbital evolution puts the effective tidal $Q_p$ of Saturn at Titan's frequency as $Q_p \approx 120$, equivalent to an evolution timescale of $a/\dot{a}=11$~Gyr \citep{Lainey-etal_2020}. Such a rapid expansion of Titan's orbit would place the origin of the Titan-Hyperion resonance at about 400-500~Myr ago.

\par The above calculation relies on two assumptions. One is that Hyperion's own rate of tidal evolution is negligible, and the other is that Hyperion did not damp its eccentricity. The first assumption about negligible tidal evolution is straightforward in the theory of equilibrium tides, in which a moon's tidal evolution rate is proportional to its mass. In the context of resonant mode locking \citep{Fuller-etal_2016}, the issue shifts from the rate of evolution to the question of whether Hyperion can lock to a mode within Saturn. Some quick calculations assuming $\dot{a}/a=11$~Gyr (i.e. the observed rate of evolution of Titan) and using standard definition of tidal parameters \citep{Murray-Dermott_1999} yield that Saturn would need to have a tidal quality factor $Q_p \simeq 10^{-3}$ at Hyperion's frequency, which is nonphysical. Therefore, Hyperion is too small and too distant to lock to a resonant mode by itself.

\par The second issue of eccentricity damping is somewhat more difficult to model as Hyperion is in chaotic rotation \citep{wis84}. One way to estimate the amount of energy dissipation is to treat it like never-ending tidal spindown \citep{Murray-Dermott_1999}. This approach gives an eccentricity damping timescale of $10^{11}$~yr (assuming $Q_H=10$ and $k_{2H}=0.01$). Alternatively, we can estimate the energy dissipation within chaotic rotation as if Hyperion was a wobbling asteroid \citep{sha05}, and this approach gives us an eccentricity damping timescale of $10^{12}$~yr. Unless another source of dissipation is identified, it appears unlikely that the eccentricity damping within Hyperion contributed significantly to the evolution of its resonance with Titan.

\par While the age of the Titan-Hyperion resonance is not the same as the age of Hyperion, there are strong reasons to think Hyperion cannot be much older than its 3:4 resonance with Titan. As Titan's and Hyperion's orbits converged, a primordial Hyperion would have likely been captured into other resonances that were crossed before the 4:3 resonance (notably the 3:2 and 7:5 resonances). Capture of Hyperion into those resonances is clearly inconsistent with Hyperion being on a relatively low-$e$, low-$i$ orbit before capture into the 4:3 resonance. While more work is needed to directly confirm the likelihood of capture into these outer resonances, we would expect at least the capture into the first-order 3:2 resonance to be robust. If Hyperion formed just interior to Titan's 3:2 resonance, Hyperion cannot be older than 1.5~Gyr, assuming Titan maintained the migration rate measured by \citet{Lainey-etal_2020}. If capture into the mutual 7:5 resonance is also found to be a certain outcome, then Hyperion must be younger than about 1~Gyr.

\par Hyperion is a relatively small moon, but its late formation would very likely imply a wider cataclysm at that time. \citet{ham13} has proposed that Titan was a late merger between multiple satellites, with Hyperion being an unaccreted fragment. \citet{asp13} also proposed a late formation mechanism for the Saturnian moons that involved a major impact on Titan. Confirming that Hyperion is not primordial would support the hypothesis of a late cataclysm involving Titan. On the other hand, if the slower migration of Titan implied by the orbital solutions of \citet{Jacobson_2022} is correct, Hyperion is likely to be primordial or almost primordial.

\subsection{Past Titan-Iapetus 5:1 MMR Crossing?}\label{iapetus}

\par Iapetus is the third-largest moon of Saturn, as well as the major moon that is the most distant from the planet. Iapetus is notable for its albedo dichotomy \citep{bur95, por05}, oblate shape \citep{tho07, cas11}, and equatorial ridge \citep{lev11, dom12, sti18}, but here we will restrict ourselves to studying its orbital motion. Like other regular satellites, Iapetus has a relatively low orbital eccentricity ($e_I=0.028$), but it also has a substantial orbital inclination ($i_I=8^{\circ}$ with respect to its Laplace plane\footnote{The instantaneous Laplace plane can be defined for every perturbed orbit as plane normal to the vector around which the orbit normal is precessing. See Section~\ref{ssec:ob-basics} for a formal definition.}), the origin of which has been a long-standing problem \citep{war81, nes14}. As the solar perturbations on Iapetus's orbit are comparable to those arising from Saturn's oblateness and the inner moons (chiefly Titan), the Laplace plane of Iapetus is significantly tilted to Saturn's equator ($i_L=14^{\circ}$). As Iapetus's orbit precesses around its Laplace plane, the instantaneous inclination of Iapetus to Saturn's equator varies approximately over a $5^{\circ}-21^{\circ}$ range over Iapetus's nodal precession period of about 3400~yr. 

\par While in the classical picture \citep[e.g.][]{Murray-Dermott_1999} Iapetus does not take part in any resonances with other satellites, faster tidal evolution \citep{lai12, lai17} would make Titan and Iapetus cross their mutual 5:1 MMR in the past. This crossing should have happened about 500 Myr ago if we assume a uniform tidal quality factor $Q=1500-2000$ for all satellites \citep{Cuk-etal_2013}, or could have happened at a very different epoch if the tidal evolution of Saturn's moons is driven by resonant modes inside the planet \citep{Fuller-etal_2016}. 

\par \citet{Cuk-etal_2013} have modeled the Titan-Iapetus 5:1 MMR crossing and found that the orbits of both bodies are chaotic during the crossing of the resonant region, which consists of numerous sub-resonances of the 5:1 resonance (they assumed $Q_p/k_{2,p}=4000$ for Saturn). If Titan was as eccentric as it is now, the most likely outcome is eccentricity growth for Iapetus, followed by orbit crossing. If the eccentricity of Titan was low (about $0.005$), Iapetus typically survives the resonance, acquiring an eccentricity of a few percent, consistent with its present orbit ($e_I=0.03$). \citet{Cuk-etal_2013}  conclude that if the Titan-Hyperion 5:1 resonance was crossed in the past (which is unavoidable for $Q_p/k_{2,p} = 4000$), the current large eccentricity of Titan must postdate this resonance crossing. This suggestion is consistent with the later excitation of Titan's eccentricity (and inclination) by an encounter with a large moon, now lost \citep{Wisdom-etal_2022} (see Section~\ref{obliquity}).

\par \citet{Cuk-etal_2013} also find that the inclination of Iapetus is affected only weakly by the 5:1 resonance crossing, with the typical change being only a degree or so, compared to the current $8^{\circ}$. Therefore, this resonance cannot constrain the timing or the source of Iapetus's inclination excitation. Interestingly, changes to the inclination of Titan can be comparable to the free inclination itself (which is $0.3^{\circ}$), making it possible that Titan's inclination was significantly modified by this resonance. 

\par \citet{Polycarpe2018} have carried out numerical simulations of the resonance crossing using an N-body code as well as using averaged equations of motion. A large span of migration rates were explored for Titan and Iapetus was started on its local Laplace plane (14$^{\circ}$ with respect to the equatorial plane) 
with a circular orbit. \citet{Polycarpe2018} find that the resonance crossing can trigger a chaotic evolution of the eccentricity and the inclination of Iapetus. The outcome of the resonance is highly dependent on the migration rate (or equivalently on $Q_p$). For a quality factor $Q_p$ of over around 2000, the chaotic evolution of Iapetus in the resonance leads in most cases to its ejection, while simulations with a quality factor between 100 and 2000 show a departure from the resonance with post-resonant eccentricities spanning from 0 up to 15\%, and free inclinations capable of reaching 11$^{\circ}$. Usually high inclinations come with high eccentricities but some simulations (less than 1\%) show elements compatible with Iapetus' current orbit. \citet{Polycarpe2018} conclude a quality factor between 100 and 2000 at the frequency of Titan would bring Titan and Iapetus into a 5:1 resonance, which could in some cases perturb Iapetus' eccentricity and inclination to values observed today. Such rapid tidal migration would have avoided Iapetus' ejection around 40-800 Myr ago.


\par \citet{Cuk2018} find that Iapetus is currently in a secular resonance with an argument $\varpi-\Omega+\varpi_J-\Omega_{eq}$ librating around 180$^{\circ}$, where $\varpi$ and $\varpi_J$ are the longitudes of pericenter of Iapetus and Jupiter, while $\Omega$ and $\Omega_{eq}$ are the longitudes of Iapetus's ascending node and  Saturn's vernal equinox. The libration period is several Myr and the libration is likely to persist for several tens of Myr. Longer-term stability of this resonance is tied to the precession of Saturn's spin axis, and more definite predictions await better determinations of Saturn's precession rate (see Section~\ref{obliquity}). Most allowable solutions for Saturn's pole precession lead to eventual breaking of the Iapetus-$g_5$ secular resonance\footnote{The eigenmode with fundamental frequency $g_5$ dominates the apsidal precession of Jupiter, enabling us to use $\varpi_J$ as its proxy.}
, but some solutions preserve the secular resonance for at least 100 Myr. The Iapetus-$g_5$ secular resonance was almost certainly established more recently than the proposed 5:1 MMR crossing between Titan and Iapetus (500-50 Myr ago, depending on the Titan's unknown tidal evolution rate). While \citet{Cuk2018} find cases when the secular resonance was established in the aftermath of this MMR (with the more rapidly evolving Titan offering promising results), they did not find a high-probability mechanism for establishing the secular resonance. 

\par There are a number of profound mysteries regarding Iapetus's orbital history. There is no agreed-upon high-probability mechanism of producing Iapetus's very large inclination. The history of its secular resonance is also unclear. The eccentricity of Iapetus is probably the orbital parameter that is most closely coupled to the overall evolution of the system. Survival of Iapetus with a low eccentricity after the 5:1 Titan-Iapetus resonance is consistent with rapid evolution found by \citet{Lainey-etal_2020}, with no restrictions on Titan's own eccentricity at the time. However, it is also consistent with a slower evolution of Titan \citep{Jacobson_2022}, but this requires that Titan acquired its eccentricity in the last 500~Myr \citep{Cuk-etal_2013}, requiring some kind of catastrophic event \citep[cf. ][]{asp13, ham13, Cuk2016}. It is still not clear whether the low eccentricity of Iapetus can be consistent with a recent loss of a moon exterior to Titan \citep{Wisdom-etal_2022}, as such a moon would strongly interact with Iapetus before its disruption.

\section{Obliquity of Saturn}\label{obliquity}

In the previous sections, specific features of the satellite dynamics were reviewed in the context of strong tidal dissipation within Saturn. On a gigayear timescale, however, satellites cannot be considered in isolation; they are part of a vast coupled system that comprises the spin-axis dynamics of their host planet. The fast migration of Saturn's satellites, and of Titan in particular, has been shown to strongly affect the motion of Saturn's spin axis (see \citealp{Saillenfest-etal_2021a,Saillenfest-etal_2021b,Saillenfest-Lari_2021,Wisdom-etal_2022}). In this section, we review the main implications of these findings.

\subsection{Basic effect of Saturn's satellites}\label{ssec:ob-basics}
\begin{figure}
   \centering
   \includegraphics[width=0.65\columnwidth]{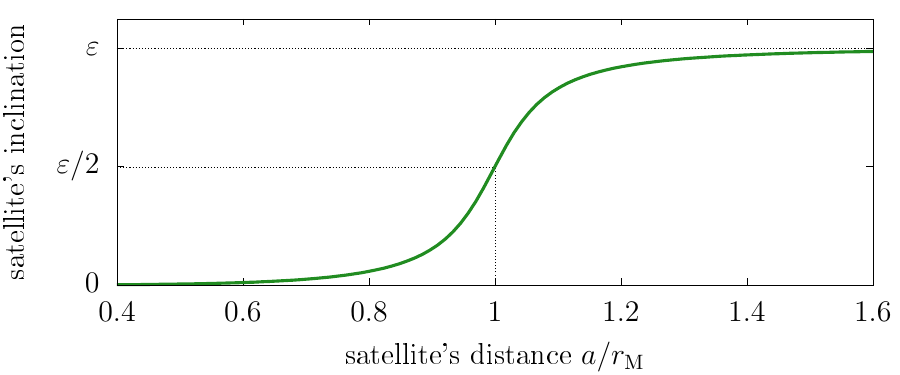}
   \caption{Equilibrium orbital inclination of a small satellite as a function of its semi-major axis. The orbital inclination of the satellite is measured with respect to the planet's equator. $\varepsilon$ is the planet's obliquity.}
   \label{fig:Lap}
\end{figure}

From a satellite's perspective, the orientation of a planet's spin axis has direct dynamical consequences. During their formation in a disc, regular moons are naturally damped towards an equilibrium configuration with near-zero eccentricity. Their orbital inclination at equilibrium defines the ``Laplace plane'', which results from a balance between the attraction of the equatorial bulge of the planet and the attraction of the star. For a massless satellite, the inclination of the Laplace plane measured from the planet's equator is:
\begin{equation}\label{eq:IL}
   I_\mathrm{L} = \frac{\pi}{2} + \frac{1}{2}\mathrm{atan2}\big[-\sin(2\varepsilon),-r_\mathrm{M}^5/a^5-\cos(2\varepsilon)\big]\,,
\end{equation}
(see \citealp{Tremaine-etal_2009,Saillenfest-Lari_2021}), where $\varepsilon$ is the planet's obliquity. As shown in Fig.~\ref{fig:Lap}, the characteristic length $r_\mathrm{M}$ is the distance at which the satellite's Laplace plane lies exactly halfway between the equator and the orbital plane of the planet (the index $\mathrm{M}$ stands for `midpoint'). It can be written
\begin{equation}
   r_\mathrm{M}^5 = 2\frac{M}{m_\odot}J_2R_\mathrm{eq}^2a_\mathrm{P}^3(1-e_\mathrm{P}^2)^{3/2}\,.
\end{equation}
Here, $m_\odot$ is the mass of the star, and $M$, $J_2$, and $R_\mathrm{eq}$ are the mass, second zonal gravity coefficient, and equatorial radius of the planet. The orbit of the planet around the star is described by its semi-major axis $a_\mathrm{P}$ and eccentricity $e_\mathrm{P}$. The value of $r_\mathrm{M}$ for Saturn is about $42$~$R_\mathrm{eq}$, so that the Laplace planes of most satellites are close to Saturn's equator. After their formation, regular moons keep oscillating about their local Laplace planes. Among the regular satellites of Saturn, only Iapetus presents today a substantial deviation from the exact equilibrium, about which it oscillates with an offset of $8^\circ$ (see Section~\ref{iapetus} of this chapter). From Eq.~\eqref{eq:IL}, we deduce that the spin-axis orientation of the planet directly sets the mean orbital inclination of its moons. Saturn's obliquity change resulting from high dissipation within Saturn (see below) has therefore an indirect effect on Saturn's whole satellite system.

From a planet's perspective, a system of satellites has a long-term contribution to its spin-axis motion. In absence of satellites, the average motion of a planet's spin axis is merely due to the stellar torque applied on its equatorial bulge. Noting $\mathbf{s}$ as the unit spin-axis vector and $\mathbf{n}$ the unit orbit normal, this motion can be written
\begin{equation}\label{eq:sdot}
   \dot{\mathbf{s}} = -\alpha (\mathbf{n}\cdot\mathbf{s})(\mathbf{n}\times\mathbf{s})\,,
\end{equation}
(see e.g. \citealp{Goldreich_1966,Tremaine_1991}), where $\mathbf{n}\cdot\mathbf{s}=\cos\varepsilon$. This equation produces a precession of the spin axis at a rate $\alpha\cos\varepsilon$ around the instantaneous orbit normal $\mathbf{n}$ of the planet. The timescale of this precession is set by the ``precession constant'' $\alpha$, which can be expressed as
\begin{equation}\label{eq:alp}
   \alpha = \frac{3}{2}\frac{G m_\odot}{a_\mathrm{P}^3(1-e_\mathrm{P}^2)^{3/2}}\frac{J_2}{\lambda\omega}\,,
\end{equation}
where $\lambda$ is the normalised polar moment of inertia of the planet, and $\omega$ is its spin rate. Satellites alter this motion in several ways: they contribute to the angular momentum of the system, they apply a torque on the planet's equatorial bulge, and their orbits are themselves torqued by the stellar attraction. The relative magnitude of these effects depends on the satellite distances. The moon-planet coupling is most obvious for the Moon around the Earth, as the lunar mass is about $1\%$ of the Earth's mass (for comparison, Titan's mass is only $2\times 10^{-4}$ of Saturn's mass). As such, the first formulas that took into account the coupled motion of a planet's spin axis and the orbit of its moons were developed for the Earth-Moon system (see e.g. \citealp{Goldreich_1966}). From these formulas, we know that satellites enhance the mean spin-axis precession rate of their host planet in a way that is intimately linked to their local Laplace plane. Analytical expressions were obtained by \cite{Ward_1975} in the two extreme cases of close-in satellites (whose mean orbital plane is the planet's equator) and far-away satellites (whose mean orbital plane is the planet's own orbital plane; see Fig.~\ref{fig:Lap}). The first closed-formed expressions valid for any distance of the satellites were introduced by \cite{Tremaine_1991} and \cite{French-etal_1993}. These expressions can be written as a modified precession constant $\alpha'$ for the planet, obtained by replacing $J_2$ and $\lambda$ in Eq.~\eqref{eq:alp} by their effective values:
\begin{equation}\label{eq:J2prime}
   \begin{aligned}
      J_2' &= J_2 + \frac{1}{2}\sum_k\frac{m_k}{M}\frac{a_k^2}{R_\mathrm{eq}^2}\frac{\sin(2\varepsilon-2I_k)}{\sin(2\varepsilon)}\,, \\
      \lambda' &= \lambda + \sum_k\frac{m_k}{M}\frac{a_k^2}{R_\mathrm{eq}^2}\frac{n_k}{\omega}\frac{\sin(\varepsilon-I_k)}{\sin\varepsilon}\,.
   \end{aligned}
\end{equation}
In this expression, the sum runs over all regular satellites (assumed to have circular orbits); $m_k$ is the mass of the $k$th satellite, $a_k$ is its semi-major axis, and $I_k$ is the inclination of its local Laplace plane with respect to the planet's equator. Equation~\eqref{eq:J2prime} can be retrieved as a particular case of the self-consistent expressions of \cite{Boue-Laskar_2006}: here, the motion of satellites is averaged over their precession around the Laplace plane. This approach is valid as long as they precess much faster than the planet's spin axis. This condition is well verified in practice for any distance of the satellites. As a result, the planet and its satellites rigidly precess as a whole \citep{Goldreich_1965}.

\begin{figure}
   \centering
   \includegraphics[width=0.65\columnwidth]{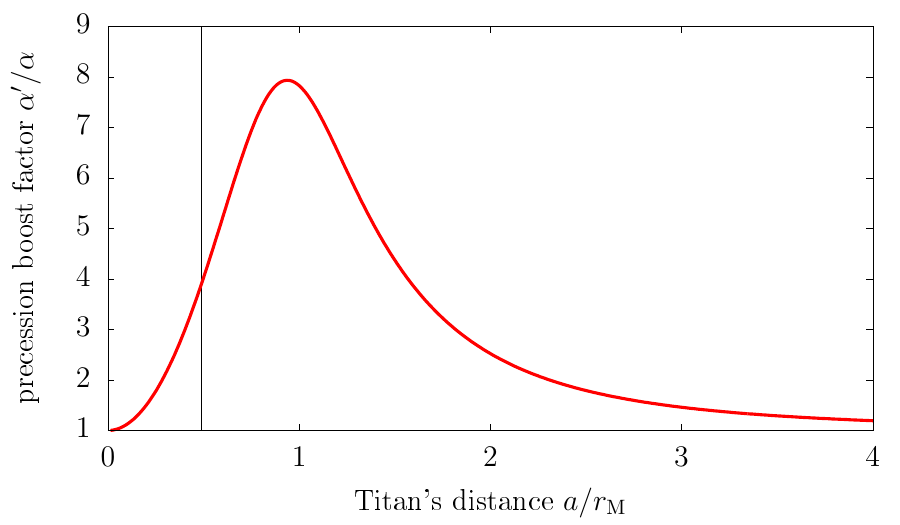}
   \caption{Enhancement factor of Saturn's spin-axis precession rate due to its most massive satellite, Titan. The current semi-major axis of Titan is shown by a vertical line. The obliquity of Saturn is fixed to its current value $\varepsilon\approx 27^\circ$. The curve reaches its maximum at $(a/r_\mathrm{M})^5\approx (\sqrt{\cos^2(2\varepsilon)+24}-\cos(2\varepsilon))/6$, see \cite{Saillenfest-Lari_2021}.}
   \label{fig:boost}
\end{figure}

Equation~\eqref{eq:J2prime} implies that the boost imparted by a moon on a planet's spin-axis precession rate strongly depends on the moon's distance. As an example, Fig.~\ref{fig:boost} shows the enhancement factor of Saturn's spin-axis precession rate as a function of Titan's distance. Despite the small mass of Titan as compared to Saturn's, the spin-axis precession rate of Saturn is currently multiplied by roughly a factor $4$ due to the presence of Titan. This precession boost strongly depends on Titan's distance. The high dissipation within Saturn and the resulting fast migration of Titan therefore have strong implications for Saturn's spin-axis dynamics.

\subsection{Secular spin-orbit resonances}
Because of the existence of secular spin-orbit resonances, a variation in a planet's spin-axis precession rate can drastically change its spin-axis dynamics. To understand this mechanism, we must notice that the orbit pole $\mathbf{n}$ appearing in Eq.~\eqref{eq:sdot} is actually not fixed, but it precesses itself as a result of mutual planetary perturbations. As the orbital motion of the planet remains largely unaffected by its spin state, the vector $\mathbf{n}$ behaves in Eq.~\eqref{eq:sdot} as an autonomous forcing term. Building on the early works by Lagrange and Laplace, very accurate theories now exist for the long-term orbital motion of the Solar System planets. As an example, Table~\ref{tab:zetashort} gives the dominant terms of the secular motion of Saturn's orbit pole computed by \cite{Laskar_1990}. The solution is written through the complex variable
\begin{equation}
   \zeta = \sin\frac{I}{2}\exp(i\Omega) = \sum_kS_k\exp[i\phi_k(t)]\,,
\end{equation}
in which the amplitudes $S_k$ are real constants and the sum runs over all terms with non-negligible amplitude. The angles $\phi_k$ evolves linearly over time as $\phi_k(t) = \nu_k\,t + \phi_k^{(0)}$, where $\nu_k$ is a fixed frequency and $\phi_k^{(0)}$ is the phase at $t=0$.

\begin{table}
   \begin{equation*}
      \begin{array}{rcrrrr}
         \hline
         \hline
         k & \text{identification} & \nu_k\ (''\,\text{yr}^{-1}) & S_k\times 10^6 & \phi_k^{(0)}\ (^\text{o})  & P_k\ (\text{kyr})\\
         \hline   
           1 &          s_5 &   0.00000 & 13774 & 107.59 &      \\
           2 &          s_6 & -26.33023 &  7850 & 127.29 &   49 \\
           3 &          s_8 &  -0.69189 &   560 &  23.96 & 1873 \\
           4 &          s_7 &  -3.00557 &   391 & 140.33 &  431 \\
           5 &  g_5-g_6+s_7 & -26.97744 &    59 &  43.05 &   48 \\
           6 &     2g_6-s_6 &  82.77163 &    34 & 128.95 &   16 \\
           7 &  g_5+g_6-s_6 &  58.80017 &    20 & 212.90 &   22 \\
           8 &     2g_5-s_6 &  34.82788 &    16 & 294.12 &   37 \\
           9 &          s_1 &  -5.61755 &    14 & 168.70 &  231 \\
          10 &          s_4 & -17.74818 &    13 & 123.28 &   73 \\
         \hline
      \end{array}
   \end{equation*}
   \caption{First ten terms in the series decomposition of Saturn's inclination and longitude of ascending node, from \cite{Laskar_1990}. The orbital elements of Saturn are measured in the J2000 ecliptic and equinox reference frame. Terms are sorted by decreasing amplitude and identified in terms of the fundamental frequencies $g_j$ and $s_j$ of the Solar System \citep[using standard definitions, see e.g.][]{Murray-Dermott_1999}. The period associated with each frequency $\nu_k$ is given in the last column.}
   \label{tab:zetashort}
\end{table}

Each term in the motion of the orbit pole $\mathbf{n}$ in Eq.~\eqref{eq:sdot} can generate a resonance for the motion of the spin axis $\mathbf{s}$. These resonances are called ``secular spin-orbit resonances'', because they involve the precession of the spin axis and the precession of the orbit. The geometry of the lowest-order resonances is given by Colombo's top Hamiltonian (see \citealp{Colombo_1966,Henrard-Murigande_1987,Saillenfest-etal_2019,Su-Lai_2022}). As a function of the planet's precession constant~$\alpha$, the system near a resonance can have up to four equilibrium points which are called the ``Cassini states'' \citep{Peale_1969}. For a critical value of $\alpha$, a separatrix appears around the Cassini state~2; this separatrix divides the inside from the outside of the resonance. At this point, it becomes clear how even a set of small moons can dramatically affect a planet's spin dynamics: because of their effective enhancement of the precession constant $\alpha'$, the presence of moons may result in the planet being inside instead of outside a given resonance. On top of that, the moons' tidal migration may slowly drive the value of $\alpha'$ past the critical value and produce a resonance capture. In case of resonance capture, the spin-axis precession rate of the planet becomes roughly constant, as it remains close to a given frequency $\nu_k$ appearing in the orbital precession (see Table~\ref{tab:zetashort}). In other words, $\alpha'\cos\varepsilon\approx -\nu_k$. As the moons continue to migrate and produce large changes in the value of $\alpha'$ (see Fig.~\ref{fig:boost}), the planet's obliquity $\varepsilon$ is forced to evolve so as to maintain the equality. This phenomenon is likely to happen for Jupiter in the future and produce an increase in its obliquity \citep{Lari-etal_2020,Saillenfest-etal_2020}.

\subsection{Saturn is close to a resonance...}\label{ssec:SaturnRes}
By taking into the account the effect of Saturn's satellites, \cite{Ward-Hamilton_2004} and \cite{Hamilton-Ward_2004} have shown that Saturn is today very close to a strong secular spin-orbit resonance with $s_8$ (third term in Table~\ref{tab:zetashort}). It was immediately recognised that this striking match and Saturn's large obliquity ($\varepsilon\approx 27^\circ$) are unlikely to be coincidental. Indeed, due to the process of gas accretion, giant gaseous planets are expected to form with a near-zero obliquity ($\varepsilon\approx 0^\circ$); some mechanism should therefore have tilted Saturn after its formation. As the strong resonance with $s_8$ has necessarily affected Saturn's obliquity in the past, it offers a natural explanation for its large obliquity (see \citealp{Ward-Hamilton_2004,Hamilton-Ward_2004}). Yet, in order to produce a resonance capture and tilting, the two frequencies involved (i.e. Saturn's spin-axis precession rate and Neptune's precession mode $s_8$) must have evolved and crossed the $1$ to $1$ commensurability. The question is when this crossing has happened in the past, and under which circumstances.

It is known today that shortly after the planetary formation, the orbits of the giant planets have been reshaped, as planets were migrating through a swarm of planetesimals \citep{Malhotra_1993}. The last large-scale changes in the orbits of the Solar System planets probably happened during this stage, more than four gigayears ago, when planetary migration triggered a phase of orbital instability (see e.g. \citealp{Tsiganis-etal_2005,Nesvorny-Morbidelli_2012}). The dominant orbital precession modes of the planets are a direct function of their semi-major axes (see e.g. \citealp{Murray-Dermott_1999}); for this reason, planetary migration and instability will have produced a variation in the frequency $s_8$. By virtue of  Eq.~\eqref{eq:alp}, Saturn's spin-axis precession rate also varied during this period of time. Hence, planetary migration first appeared to astronomers as a natural explanation for the crossing of Saturn's secular spin-orbit resonance with $s_8$. More than that, the current obliquity of Saturn was viewed as a practical constraint for the planetary migration process itself \citep{Boue-etal_2009,Vokrouhlicky-Nesvorny_2015,Brasser-Lee_2015}.

However, the high tidal dissipation inside Saturn measured by \cite{Lainey-etal_2020} has changed our understanding of Saturn's tilting process. The migration rate measured for Titan is so fast that Titan has most probably migrated over a distance of several radii of Saturn in the past (and possibly even more than that; see Chapter~3 of this book, but also the contrary opinion of \citealp{Jacobson_2022}). Due to the steepness of its enhancement factor on Saturn's precession constant (see Fig.~\ref{fig:boost}), this migration for Titan must have greatly altered Saturn's precession rate over time. As a result, it appears very unlikely that Saturn crossed the $s_8$ resonance during late planetary migration, more than four gigayears ago, because its spin-axis precession motion was too slow at that time. Instead, Titan's orbital expansion offers a new explanation for Saturn's capture into resonance and tilting to its current $27^\circ$ obliquity \citep{Saillenfest-etal_2021a,Wisdom-etal_2022}.

\begin{figure}
   \centering
   \includegraphics[width=\columnwidth]{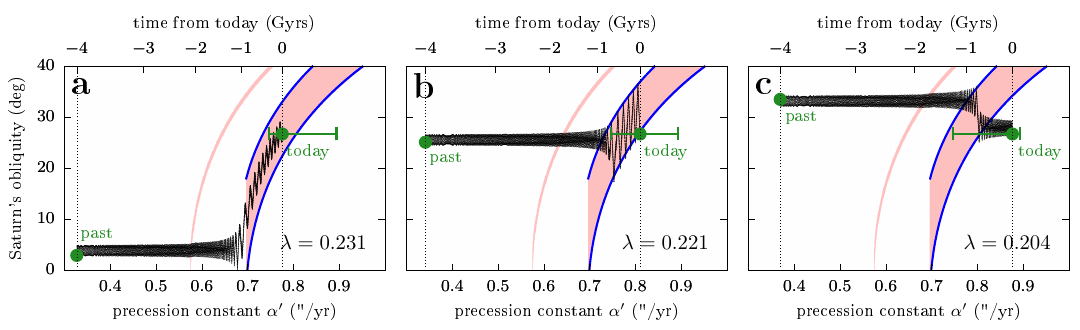}
   \caption{The three types of possible obliquity evolution for Saturn (adapted from \citealp{Saillenfest-etal_2021a}). Each panel shows an example of simulation for a given value of $\lambda$ (see labels). The explored range of values lies in the horizontal green bar. The first and last point of each trajectory are shown by a green spot. First-order secular spin-orbit resonances are shown in pink: the large area on the right is the resonance with $s_8$ and the thin area on the left is the resonance with $g_7-g_8+s_7$ (not shown in Table~\ref{tab:zetashort}; 19th term). The separatrices of the $s_8$ resonance are highlighted in blue. Panels a and b show resonant capture while panel c does not.}
   \label{fig:types}
\end{figure}

\begin{figure}
   \centering
   \includegraphics[width=0.6\columnwidth]{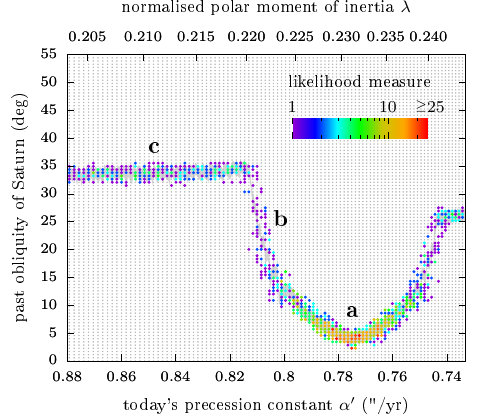}
   \caption{Likelihood of reproducing Saturn's current state as a function of its past obliquity (adapted from \citealp{Saillenfest-etal_2021b}). Each point is made of $240$ numerical simulations with random initial precession angle for Saturn, started $4$~Gyrs ago and propagated until today. The colour scale measures the relative likelihood of obtaining a final obliquity and precession angle within $0.5^\circ$ and $5^\circ$ of Saturn's current values, respectively. A small grey dot means that no successful trajectory was found. The labels a, b, and c refer to the trajectory types illustrated in Fig.~\ref{fig:types}.}
   \label{fig:CIsearch}
\end{figure}

The actual dynamical path followed by Saturn depends on its current location with respect to the resonance. As described by Eq.~\eqref{eq:alp} and \eqref{eq:J2prime}, the precise precession rate of Saturn depends on several physical parameters. Each of these parameters are well known today from direct measurements, except Saturn's normalised polar moment of inertia $\lambda$. Indeed, the gravitational potential measured by spacecraft only provides differences between the moments of inertia ($J_2$, $J_4$, etc.), but no measure of the individual moments of inertia. In order to obtain the value of Saturn's polar moment of inertia, authors usually use semi-empirical models of Saturn's interior structure that are fitted to Saturn's gravitational moments (see e.g. \citealp{Hubbard-Marley_1989,Nettelmann-etal_2013,Movshovitz-etal_2020}). Such estimates, although formally accurate, are model-dependent and generally do not agree with each other. Moreover, it is not clear how to convert the moment of inertia derived from state-of-the-art interior models to that used in simplified models of rotation dynamics (which use the gyroscopic approximation, a purely rigid rotation, no internal dynamical processes, etc.); the small discrepancy between these two concepts of moment of inertia may lead to a slight shift of Saturn's precession constant in the neighbourhood of the $s_8$ secular spin-orbit resonance. For these reasons, previous authors did not rely on a given value of $\lambda$, but they explored a range of possible values spanned by the estimates obtained through interior modelling. Using this approach, \cite{Saillenfest-etal_2021a} showed that Saturn could have followed three possible types of trajectories as a result of Titan's fast migration (see Fig.~\ref{fig:types}). For a given range of $\lambda$ values, one obtains very small past obliquities for Saturn (panel~a) and a resonance capture with $100\%$ certainty without crossing the separatrix. For slightly lower values of $\lambda$, one obtains a recent resonance capture by crossing the separatrix (panel~b) or a resonance crossing without capture (panel~c). By conducting large-scale Monte Carlo experiments, \cite{Saillenfest-etal_2021b} investigated the likelihood for each pathway as a function of Saturn's initial obliquity. The result is shown in Fig.~\ref{fig:CIsearch} for Titan's nominal migration rate measured by \cite{Lainey-etal_2020}. Trajectories labelled~``b'' are by far the less likely, because separatrix crossings are probabilistic events and in this case they rarely lead to a capture (see \citealp{Saillenfest-etal_2021b}). Trajectories labelled~``a'' are about ten times more likely that those labelled~``c''. This higher likelihood is mostly due to the precession phase of Saturn's spin axis, which points today roughly in the direction of the centre of the resonance\footnote{\label{ftn:sigma}The current value of the resonance angle is $\sigma\approx 30^\circ$, which is relatively close to zero \citep{Ward-Hamilton_2004}. If Saturn is not locked in resonance today (as in case~c), then $\sigma$ would circulate between $0^\circ$ and $360^\circ$ and have no particular reason to be small today; it would rather be a coincidence. If instead Saturn is locked in resonance today (as in case~a), then $\sigma$ would oscillate around $0^\circ$ and always keep a relatively small value.}. Hence, the tilting of Saturn from a small obliquity -- and Saturn still being in resonance today -- is by far the most likely scenario from a dynamical point of view.

In this situation, Saturn's obliquity would continue to increase in the future as a result of the still ongoing migration of Titan; it could even reach values as high as $\varepsilon\gtrsim 75^\circ$ within $5$~Gyrs, at which point Titan would be strongly destabilised (see \citealp{Saillenfest-Lari_2021}).

\subsection{...but Saturn is just out of sync}
The dynamical pathway followed by Saturn depends on its current location with respect to the $s_8$ resonance. If Saturn is inside the resonance today, then it would have been captured from a small obliquity and it would continue to follow the resonance as Titan migrates (Fig.~\ref{fig:types}a). If instead Saturn is outside the resonance today, then it would have recently crossed the resonance without capture, from a higher obliquity (Fig.~\ref{fig:types}c). In order to determine whether Saturn is currently inside or outside the resonance, \cite{Wisdom-etal_2022} developed new models of Saturn's interior structure that were fitted to the latest estimates of Saturn's gravitational moments measured by the Cassini spacecraft. From these models, they deduced refined values of Saturn's polar moment of inertia $\lambda$ (estimated to be equal to $\rm 0.2182^{+0.0006}_{-0.0003}$) that they used in long-term numerical integrations of Saturn's spin axis.

According to their results, Saturn is not in resonance today, but just outside its edge, at the level of $1\%$ on the value of $\lambda$. As for the ambiguity and model dependency inherent to this approach (see above), \cite{Wisdom-etal_2022} argue that the four models that they consider cover the entirety of possible solutions that one could obtain from realistic models of Saturn's interior. The more recent interior models of \citet{Mankovich-etal_2023}, which include a fit to ring seismology, also suggest that Saturn is out of resonance at present (see below). On this basis, Saturn should definitely be considered out of the $s_8$ resonance today. Saturn should therefore follow a pathway similar to that in Fig.~\ref{fig:types}c.

\begin{figure}
   \centering
   \includegraphics[width=\columnwidth]{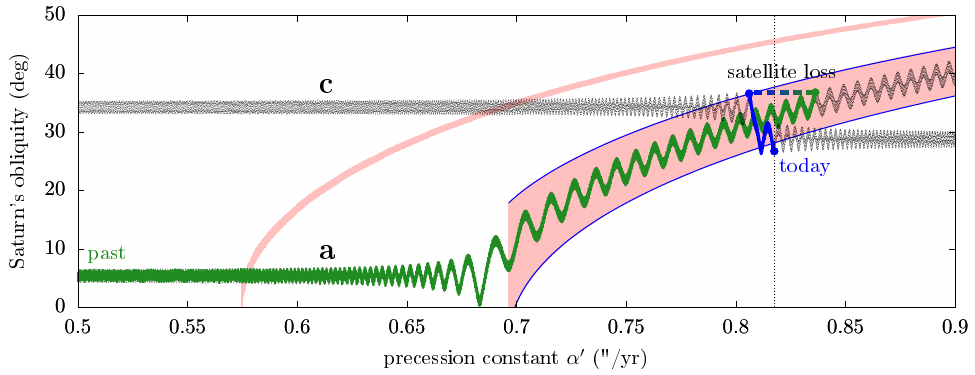}
   \caption{Variant scenario proposed by \cite{Wisdom-etal_2022}. The small black dots show the types of trajectories~``a'' and ``c'' depicted in Fig.~\ref{fig:types}. The two portions of pathway followed by Saturn are superimposed in green and blue, respectively. When Saturn's ancient satellite is lost, the curve instantly jumps from the green to the blue curve (dashed segment).}
   \label{fig:W22}
\end{figure}

In this case, we are back to the question of the origin of Saturn's large obliquity. As recalled by \cite{Saillenfest-etal_2021a}, an early giant impact could be a possibility. Instead of a giant impact, \cite{Wisdom-etal_2022} propose to connect the two possible types of pathways (``a'' and ``c'') by a timely reorganisation in Saturn's moon system. Indeed, if a moon of Saturn is suddenly ejected or engulfed into Saturn, then Saturn's effective precession constant $\alpha'$ would jump to a lower value (see Eq.~\ref{eq:J2prime}). If this moon is massive enough, this would instantly kick Saturn out of the resonance. \cite{Wisdom-etal_2022} show that a relatively small ancient moon would be enough to produce this effect, for instance with a mass of the order of Iapetus's. Then, the still ongoing migration of Titan would make Saturn cross the resonance again, but without capture. This scenario is illustrated in Fig.~\ref{fig:W22}. Ejection of the moon is hypothesized to occur when its orbit is destabilized during a 3:1 MMR with Titan \citep{Wisdom-etal_2022}.

As Saturn is still very close to the resonance today, and Titan's migration drives Saturn farther and farther away as time goes by, the removal of Saturn's ancient moon must have occurred recently in the history of the Solar System. For a former moon with the mass of Iapetus (that would not kick Saturn very far off the resonance), \cite{Wisdom-etal_2022} estimate that this event should have happened about $100$~Myrs ago. Such a recent event reminds us of the controversial young age of Saturn's ring (see Chapter~11 of this book). \cite{Wisdom-etal_2022} thus suggest to link these two phenomena: the lost moon would have been ripped apart within Saturn's Roche limit, which would have supplied the ring material, similarly to what has been proposed by \cite{Canup_2010}. A recent instability in Saturn's system may also explain some peculiarities in the moons' orbits, such as Titan's eccentricity or Iapetus's inclination (see Section~\ref{iapetus}). It is still unclear, however, how such a catastrophic event would relate to the existence of Saturn's mid-sized moons, which all have rather unexcited orbits today and are connected through a fragile web of MMR (see Section~\ref{passages}).


The scenario of \cite{Wisdom-etal_2022} has the merit of preserving Titan's migration as the explanation for Saturn's large obliquity (see \citealp{Saillenfest-etal_2021a}), while allowing Saturn to be out of the resonance today. However, the likelihood analysis shows that, if one remains somewhat sceptical about the exact value of Saturn's polar moment of inertia, Saturn's current dynamical state is about ten times more likely to result from Saturn still being in resonance today (mostly due to the current small value of the resonance angle $\sigma$; see footnote~\ref{ftn:sigma} above). It is also more probable in a cosmogonic point of view: if Saturn had freshly crossed the resonance, then humanity would be witnessing today a transient event that could have happened any time over the history of the Solar System. That being said, Saturn's ring does seem exceptional in the Solar System, so one may need to accept some exceptional level of chance to account for its existence (see the discussion by \citealp{ElMoutamid_2022}). Furthermore, the Saturn system seems to have possessed a population of planetocentric impactors which are not observed elsewhere \citep{ferguson2022regional} and suggest that something unusual happened. 

The debate is revived by the recent article by \cite{Jacobson_2022}. His self-consistent adjustment of Saturn's spin axis and satellite dynamics to all available data yields a lower migration rate for Titan than previously measured (see Chapter~3 of this book). If we extrapolate the results of \cite{Jacobson_2022} into the past, we find that Titan would have migrated just the right amount to explain Hyperion's eccentricity (\citealp{Cuk-etal_2013}; see Section~\ref{hyperion}) and Saturn's tilt (\citealp{Saillenfest-etal_2021b}; see Section~\ref{ssec:SaturnRes}) over the age of the Solar System. Moreover, \cite{Jacobson_2022} obtains a value for Saturn's moment of inertia\footnote{When converted to the same normalising radius $R_\mathrm{eq}$ and spin rate $\omega$ as those used in Section~\ref{ssec:SaturnRes}, the values obtained by \cite{Jacobson_2022} for Saturn's normalised moment of inertia are $\lambda=0.2283$ and $0.2313$ (see Eq.~\ref{eq:alp}), with a $1\sigma$ uncertainty of about $1\%$. The two values correspond to two different sets of observations.} that would put Saturn near the centre of the most probable region in Fig.~\ref{fig:CIsearch}; this would imply that Saturn's obliquity has evolved as in Fig.~\ref{fig:types}a, rather than the scenario proposed by \cite{Wisdom-etal_2022} and displayed in Fig.~\ref{fig:W22}. If so, then the origin of Saturn's rings, Titan's eccentricity and the population of planetocentric impactors all remain to be explained. In the next few years, we can expect different teams of astronomers to revisit the studies of \cite{Wisdom-etal_2022} and \cite{Jacobson_2022}; the future will tell which scenario will reach scientific consensus. Similarly, future work will hopefully yield agreement on the correct value for Saturn's polar moment of inertia $\lambda$.

A recent step in this direction has been made by \cite{Mankovich-etal_2023}. The authors fitted interior models not only to the measured gravity harmonics of Saturn, but also on the frequencies of Saturn's internal oscillation modes imprinted in the rings. This approach allows the deep rotation profile of Saturn to be probed. From their models of internal structure, \cite{Mankovich-etal_2023} computed the expected value of Saturn's moment of inertia: they note that their results are consistent with the upper end of estimates of \cite{Wisdom-etal_2022}, and approximately $0.5\%$ smaller than the value required for the $s_8$ resonance to be active today. This supports the notion that Saturn is today outside the resonance, though close to its edge. \cite{Mankovich-etal_2023} also stress that large residuals in the frequencies of several modes (mostly for Saturn's deep structure, but also for outer layers) reflect the fact that current models for Saturn's interior are still not complete. According to them, a parallel can be drawn between this mismatch and the persistent difficulty encountered in fitting Jupiter and Saturn interior models to gravity while retaining the supersolar envelope heavy element enrichments indicated by spectroscopic measurements. These discussions reveal the richness and complexity of planetary interior modelling, and new avenues to better match the available observational constraints. \cite{Mankovich-etal_2023} advocate giving more freedom to the composition, temperature, and rotation profiles used as working interior models for their adjustments. It will be interesting to know whether such future refined interior models could change estimates for Saturn's moment of inertia by a non-negligible amount, and whether the strongly non-rigid rotation of Saturn that they reveal -- still never taken into account in simulations of Saturn's long-term spin-axis dynamics -- can alter in some way the precession rate of Saturn's spin axis.

\section{Interior evolution of moons}\label{interior}


The interior evolution of moons is driven by heating processes that can affect the state (e.g., solid or liquid), location (e.g., homogeneous or segregated), and composition of their materials. These materials appear to be primarily ices (chiefly water ice) and, as inferred from meteorites and interplanetary dust particles, silicate rock and refractory organic material. The proportion of ice is inferred from the moons' bulk densities (985 kg m$^{-3}$ for Tethys to 1879 kg m$^{-3}$ for Titan) to be at least 50 vol.\%, with some uncertainty arising from porosity (for smaller moons such as Mimas) and compression (for Titan). Heating predominantly arises from the long-term decay of radioactive isotopes sourced in silicate rock and from tidal dissipation (Eq. \ref{eq:Edot}) with additional contributions, likely limited in time, from the radioactivity of short-lived isotopes, release of potential energy during accretion or separation of materials by density (differentiation), and chemical reactions between materials (e.g., rock hydration). Radiogenic heating drives larger, more silicate-rich moons, with a lower ratio of surface area (heat loss) to volume (heat production), to have warmer, more evolved interiors. Tidal heating drives closer-in, more eccentric or inclined moons to be warmer and thus more evolved (Eq. \ref{eq:Edot}). Tidal dissipation also couples the interior and orbital evolution of a moon and, because of gravitational interactions (e.g., resonances) within the moon system, to those of the other moons, Saturn, and its rings as well. While interior and orbital evolution aspects can and have been studied separately for specific satellites, or a subset of the moon system, system-level studies of the coupled interior-orbital evolution are challenging to carry out owing to the diversity of timescales involved (from orbital periods of days to radioactive decay timescales of billions of years) and the larger space of uncertain or unknown parameters.

A number of observational constraints provide a starting point. In addition to the dynamical constraints on ongoing and past orbital evolution described in previous sections, these include geological observations (see Chapter 5 for an in-depth discussion). The moons' ice-rich surfaces and (assuming hydrostatic equilibrium) moments of inertia indicate some degree of ice-rock separation, but this separation is incomplete inside at least Titan \citep{iess2010gravity} and perhaps Enceladus \citep{hemingway2019enceladus} and Rhea \citep{tortora2016rhea}. 
The data at other moons are too uncertain to tell \citep{tajeddine2014constraints, beuthe2016enceladus}. Enceladus (see Chapter 5) and, perhaps, Titan \citep{nixon2018titan} show evidence of ongoing endogenic geological activity, and Tethys and Dione show evidence of past activity linked to past high heat fluxes \citep{ChenNimmo2008,hammond2013flexure,white2017impact}. The sizes of the moons increase with increasing semi-major axis from Mimas to Titan, but their bulk densities (rock content, including heat-producing radioisotopes) do not.

An especially noteworthy observation, the `Mimas paradox', is the puzzling contrast in geological activity between inactive Mimas and (cryo)volcanic Enceladus. From Eq. \eqref{eq:Edot}, the dissipation of solid tides raised by Saturn, which dominates radiogenic heating at these small, close-in moons, is roughly 40 times stronger at Mimas than at Enceladus, assuming a similar interior structure and material response (i.e., $k_2/Q$). Why, then, is Enceladus so active and Mimas so quiescent? Proposed solutions to this paradox involve different interior material responses between Mimas and Enceladus, i.e., different loci in the positive feedback loop in which increasing interior temperatures lead to decreasing material viscosities, which lead to increasing tidal dissipation and more heating. \citet{malamud2013modeling} proposed that Enceladus was nudged into this positive feedback loop by a spike of chemical heating during the hydration of the core, possible at Enceladus but not Mimas due to the former's higher radiogenic heat supply (Enceladus is denser and thus has a higher rock fraction than Mimas). \citet{czechowski2015comparison} suggested that this nudging could have also been caused by higher short-lived radiogenic heating at Enceladus. \citet{neveu2019evolution} invoked more radical differences: a young age for Mimas and earlier encounter of MMR by Enceladus which would not be today in equilibrium heating. \citet{rhoden2022case} suggested that Mimas may have an ocean just like Enceladus, which could be compatible with observations of its libration \citep{tajeddine2014constraints}. Whether this suggestion is compatible with the lack of obvious indications of extensive relaxation of Mimas's large impact basins remains to be established. An ocean-bearing Mimas would also have to have avoided damping of Mimas's eccentricity (Section~\ref{dione}).


The long-term evolution of the moons depends on how and when they formed. This is largely an open question for the Saturn system. While Titan and Iapetus are likely primordial, the age of Rhea and closer-in moons is unclear (Chapter 5). They too could have formed concurrently with Saturn at the dawn of the Solar System \citep[and references therein]{mosqueira2010planetesimals}, possibly coexisting and interacting with a long-lived massive ring \citep{nakajima2019orbital}, or from the reaccretion of moon fragments from prior collisions with Sun-orbiting impactors \citep{movshovitz2015disruption}, or from accretion at the outer edge of Saturn's rings followed by tidally driven orbital expansion \citep{charnoz2010recent}. Each formation scenario is compatible only with a subset of possible orbital expansion histories (Section~\ref{planet}): roughly, the end-members are accelerating expansion for primordial or old moons and decelerating expansion for young moons. Astrometric observations of fast orbital expansion \citep{Lainey-etal_2020} currently lean toward the former (but see \citealp{Jacobson_2022}). If moons formed from rings, rock (silicates and organics) and ice may have been separated inside moons at the time of formation: the lesser propensity of rock to undergo tidal disruption would have led first to the formation of a rock-rich core, onto which ice would then have accreted as tidally driven orbital expansion led to decreased tidal forcing \citep{charnoz2011accretion}. This appears to be the case for Mimas, whose predicted fast orbital expansion, dominated by interactions with Saturn's nearby rings, suggests a young age if Mimas postdates the rings (Chapter 5), and whose libration necessitates some degree of internal differentiation \citep{tajeddine2014constraints}. However, there is no independent evidence that Mimas postdates the rings. The higher propensity of close-in moons (Enceladus, Tethys, Dione)  to encounter MMR implies a greater potential for interior evolution via episodes of high heat flow or collisions, especially if the moons are old. This is reflected in the broader diversity of geological features at their surfaces relative to Mimas, which may be younger, and Rhea, which is farther out. 

In contrast, the interior evolution of Rhea, Titan, and Iapetus appears dominated by radiogenic heating and thus less tumultuous (see Chapter 5). Dating events in the moons' history, such as past periods of geological activity, can constrain or test hypotheses on their internal evolution \citep{Zahnle-etal:2003}; progress on crater-based dating accounting for planetocentric impactor populations \citep{ferguson2020small, ferguson2022regional} and future in situ measurements could decrease the currently considerable uncertainties on surface ages.

\section{Conclusions}\label{conclusion}

The long-term evolution of the Saturnian system has probably never been more contentious. The proposed views of the system have changed over the years, from a primordial one driven by relatively slow tides \citep{Murray-Dermott_1999, MeyerWisdo:2008}, through a fast-evolving system \citep{lai12} that may not be primordial \citep{Cuk2016}, to a resonance-lock driven one that could either be primordial \citep{Fuller-etal_2016, Lainey-etal_2020} or recently reset \citep{Wisdom-etal_2022}. Our models of the system's long-term evolution are closely related to the observational results on the Saturnian rings, satellite astrometry and crater statistics which are covered in other chapters. Some of the past controversies have been resolved, as we now know that the rings have a relatively low mass \citep{ies19}, even if their age is still in question \citep{cri19}, and the impact craters on the satellites are at least partially caused by planetocentric debris \citep{ferguson2020small, ferguson2022regional}. New controversies include the current migration rate of Titan \citep{Lainey-etal_2020, Jacobson_2022} and whether Saturn's spin precession is still in a resonance with Neptune's orbit \citep{Saillenfest-etal_2021a, Wisdom-etal_2022}. 

One firm data point is that the observed tidal evolution of Rhea is well constrained \citep{lai17, Lainey-etal_2020, Jacobson_2022} and is incompatible with equilibrium tides, indicating that highly frequency-dependent (probably resonant) tides must be operating in the Saturnian system (Chapter 3). While this is technically just an instantaneous migration rate, the existence of resonant tides greatly influences our understanding of the past evolution of the system. In this chapter we reviewed past work on the long-term dynamical evolution of Saturn's satellites, and attempted to show how various resonances and other dynamical features constrain the current theories about the system's age and tidal evolution. Our findings can be summarized as follows:

1. The currently observed Mimas-Tethys 4:2 and Enceladus-Dione 2:1 resonances indicate that in each of these two pairs, the satellites are on converging orbits. This is incompatible with all satellites being on diverging \citep{Fuller-etal_2016} or parallelly-evolving orbits \citep{Lainey-etal_2020}. If the system is primarily driven by resonance locking, these resonances indicate that not all satellites are currently locked to Saturn's internal modes.

2. The current heat flux of Enceladus (if it is in equilibrium) is consistent both with strong equilibrium tides \citep{lai12} or diverging resonant modes \citep{Fuller-etal_2016}, but is likely significantly smaller than the expected tidal heating if Enceladus is locked to one of the parallelly-evolving modes \citep[as illustrated in Fig. 3 of][]{Lainey-etal_2020}. 

3. The excitation of Tethys's 1$^{\circ}$ inclination is an important constraint on the evolution of the system before the current resonances. It is possible that Tethys's inclination is a consequence of Dione-Rhea 5:3 resonance crossing followed by Tethys-Dione secular resonance \citep{Cuk2016}, but that scenario would require the current fast evolution of Rhea to be a very recent phenomenon. If Rhea's current rate of evolution is a long-term value, then Dione and Rhea never crossed their 5:3 resonance. 

4. Another ``fossil'' feature of the inner system is the excited eccentricity of Mimas, which could have been excited by the past crossing of the Mimas-Dione 3:1 resonance \citep{Cuk_ElMoutamid_2022}.

5. In every modern theory of its tidal evolution, Rhea should have crossed the semi-secular evection resonance with the Sun which should have excited Rhea's eccentricity and inclination well beyond the observed values. The damping of Rhea's inclination would be difficult to achieve, possibly indicating that Rhea may not be primordial but that it could have reaccreted close to its present orbit.

6. The current Titan-Hyperion 4:3 resonance has been evolved by tides and, depending on which estimate of Titan's tidal evolution rate we use, may be either 500 Myr old \citep[using][]{Lainey-etal_2020} or 4 Gyr old \citep[using][]{Jacobson_2022}. In the former case Hyperion is not primordial and must be a product of a cataclysm of some kind within the last 1-1.5 Gyr.

7. Titan and Iapetus must have crossed their mutual 5:1 resonance at some point in the past. Once again, different estimates of Titan's migration rate place this crossing at either 50 Myr ago \citep{Lainey-etal_2020} or 500 Myr ago \citep{Jacobson_2022}. The latter case of slower evolution through the resonance leads to more dynamical excitation, so in this case Titan must have acquired its eccentricity after the 5:1 resonance with Iapetus, again indicating some kind of possibly catastrophic, recent event.

8. Titan's relatively high eccentricity and mechanically weak interior have long been thought to be problematic, indicating that the eccentricity had to be generated relative recently or it would have been damped. Mechanisms for this excitation usually involve some sort of catastrophe in which at least some moons are disrupted \citep{asp13, ham13, Cuk2016, Wisdom-etal_2022}

9. A resonance between the precession of Saturn's spin axis and the orbit of Neptune almost certainly generated Saturn's obliquity \citep{Ward-Hamilton_2004, Hamilton-Ward_2004, Saillenfest-etal_2021a, Saillenfest-etal_2021b}. The substantial orbital expansion of Titan offers a natural origin for this resonance capture. \citet{Wisdom-etal_2022} find that Saturn is currently not in this spin-orbit resonance, but this claim depends on the value of Saturn's moment of inertia derived from interior models, which are not all in agreement. However, if the resonance is broken, this would strongly indicate a recent cataclysm in the outer part of the satellite system (beyond Titan), although the viability of the specific scenario of \cite{Wisdom-etal_2022} and the connection to the formation of the rings may need further confirmation. 

The above points demonstrate that there are no easy answers to the questions posed by the Saturnian system. It appears that there is no simple unified explanation for the orbital evolution of all moons (past and observed), and while some moons may not be primordial, the age and extent of any disruption events is highly uncertain. 

Fortunately, there are two observable quantities that may be constrained enough in the near future to help us narrow down the range of the system's possible histories. One is the precession rate of Saturn's spin axis. Current measured values lean toward Saturn being in the spin-orbit resonance with Neptune \citep{Jacobson_2022}, though non-resonant values are also permitted.. If Saturn is determined to be in the spin-orbit resonance, that would remove the need for (but not fully rule out) the proposed cataclysm in the outer part of the system (which is more important for Saturn's precession than the inner satellites). The second observable quantity is the current rate of tidal evolution of Titan. At the time of writing there are two contradictory results in the literature \citep{Lainey-etal_2020, Jacobson_2022} that are largely based on the same data. If a consensus could be reached on this issue, we would have better understanding of the physics of tidal evolution in the system. Titan is the largest moon and is relatively far from Saturn. As such it is most likely to be ancient and most likely to be locked to a resonant mode within Saturn. Confirmation of a fast-evolving Titan would demonstrate the reality of resonant locking; a slower migrating Titan could indicate that fast migration driven by resonant modes may not be permanent. 

Apart from the two above mentioned observable quantities, progress may come from sources of data other than the satellite dynamics. Better understanding of Saturn's interior \citep{man21}, the rings \citep{ies19} and satellite surfaces \citep[][also Chapter 5 of this book]{ferguson2022regional} could all help us understand the history of the system. Furthermore, more theoretical work on the moons' past orbital evolution, while not assumption-free, could still produce valuable new insights on the possible histories of the system, especially if combined with constraints from other lines of evidence.

\backmatter

\bmhead{Acknowledgments}

The authors thank the International Space Science Institute and Dr. Valery Lainey for organizing the workshop ``New Vision of the Saturnian System in the Context of a Highly Dissipative Saturn'' that inspired this edited volume.

\section*{Declarations}


\begin{itemize}
\item Funding

M\'C is supported by NASA Solar System Workings award 80NSSC22K0979. ME and M\'C are supported by NASA-SSW award 80NSSC23K0681. FN is supported by NASA-SSW award 80NSSC21K1802. MN is supported by NASA-CDAP award 80NSSC21K0531 and by the CRESST II agreement between NASA GSFC and Univ. Maryland, College Park (award number 80GSFC17M0002). AR is supported by NASA Cassini Data Analysis Program award 80NSSC21K0535.

\item Conflict of interest/Competing interests 

The authors have no relevant financial or non-financial interests to disclose.

\item Ethics approval 

Not applicable.

\item Consent to participate

Not applicable. 

\item Consent for publication

Not applicable.

\item Availability of data and materials

Not applicable.

\item Code availability 

Not applicable.

\item Authors' contributions

All authors have contributed to this review chapter.

\end{itemize}








\bibliographystyle{plainnat}
\bibliography{sn-bibliography} 

\end{document}